\documentclass[fleqn,usenatbib]{mnras}

\usepackage{newtxtext,newtxmath}

\usepackage{ulem}       

\usepackage{array}      
\usepackage{makecell}   

\usepackage[T1]{fontenc}

\DeclareRobustCommand{\VAN}[3]{#2}
\let\VANthebibliography\thebibliography
\def\thebibliography{\DeclareRobustCommand{\VAN}[3]{##3}\VANthebibliography}

\def\nustar{{\it NuSTAR}}
\def\nicer{{\it NICER}}

\def\maxi{{\it MAXI}}

\def\xillver{{\tt xillver}}
\def\xillverD{{\tt xillverD}}
\def\xillverCp{{\tt xillverCp}}
\def\xillverDCp{{\tt xillverDCp}}

\def\relflionx{{\tt reflionx}}

\def\reltrans{{\sc reltrans}}

\def\reltransCp{{\sc reltransCp}}
\def\reltransDCp{{\sc reltransDCp}}
\def\kynrefrev{{\sc kynrefrev}}
\def\nthcomp{{\tt nthcomp}}
\def\xspec{{\tt xspec}}
\def\logXi{$\log(\xi/$erg\,cm\,s$^{-1})$}

\def\maxi{MAXI~J1820+070}

\usepackage{graphicx}	
\usepackage{amsmath}	

\title[RELTRANS 2.0]{Modelling correlated variability in accreting black holes: the effect of high density and variable ionisation on reverberation lags}

\author[Mastroserio et al.]{
Guglielmo Mastroserio$^{1}$\thanks{E-mail: gullik@caltech.edu}, Adam Ingram$^{2}$, Jingyi Wang$^3$, Javier A. Garc{\'\i}a$^{1,6}$, Michiel van der Klis$^{4}$,
\newauthor Yuri Cavecchi$^{5}$, Riley Connors$^{1}$,   
Thomas Dauser$^{6}$, Fiona Harrison$^{1}$, Erin Kara$^{3}$, Ole K\"{o}nig$^{6}$, Matteo Lucchini$^{3}$
\\
$^{1}$Cahill Center for Astronomy and Astrophysics, California Institute of Technology, Pasadena, CA 91125, USA\\ 
$^{2}$Department of Physics, Astrophysics, University of Oxford, Denys Wilkinson Building, Keble Road, Oxford OX1 3RH, UK\\
$^{3}$MIT Kavli Institute for Astrophysics and Space
Research, MIT, 70 Vassar Street, Cambridge, MA 02139, USA\\
$^{4}$Astronomical Institute, Anton Pannekoek, University of Amsterdam, Science Park 904, NL-1098 XH Amsterdam, Netherlands\\
$^{5}$Instituto de Astronomía, Universidad Nacional Autónoma de México, Ciudad de México, CDMX 04510, Mexico\\
$^{6}$Dr Karl Remeis-Observatory and Erlangen Centre for Astroparticle Physics, Universit\"{a}t
Erlangen-N\"{u}rnberg, Sternwartstr.~7, D-96049 Bamberg, Germany\\ 
}

\date{Accepted XXX. Received YYY; in original form ZZZ}

\pubyear{2021}

\begin{document}
\label{firstpage}
\pagerange{\pageref{firstpage}--\pageref{lastpage}}
\maketitle

\begin{abstract}
We present a new release of the \reltrans\ model to fit the complex cross-spectrum of accreting black holes as a function of energy. The model accounts for continuum lags and reverberation lags self-consistently in order to consider the widest possible range of X-ray variability timescales.
We introduce a more self-consistent treatment of the reverberation lags, accounting for how the time variations of the illuminating flux change the ionisation level of the accretion disc. This process varies the shape of the reflection spectrum in time causing an additional source of lags besides the light crossing delay. 
We also consider electron densities in the accretion disc up to $10^{20}$\,cm$^{-3}$, which are found in most of the stellar mass black holes and in some AGN. These high densities increase the amplitude of the reverberation lags below $1$ keV since the reflection flux enhances in the same energy range. 
In addition, we investigate the properties of hard lags produced by variations in the power-law index of the continuum spectrum, which can be interpreted as due to roughly $3\%$ variability in the corona's optical depth and temperature. 
As a test case, we simultaneously fit the lag energy spectra in a wide range of Fourier frequency for the black hole candidate \maxi1820 observed with \nicer . The best fit shows how the reverberation lags contribute even at the longer timescales where the hard lags are important. 
This proves the importance of modelling these two lags together and self-consistently in order to constrain the parameters of the system.

\end{abstract}

\begin{keywords}
black hole physics -- relativistic processes -- accretion, accretion discs -- X-rays: binaries -- X-rays: galaxies
\end{keywords}



\section{Introduction}
Many stellar-mass black holes and supermassive black holes accrete material through a geometrically thin and optically thick disc. 
The thermal radiation of the accretion disc (`seed' photons) is thought to be inverse Compton up-scattered
by a region of hot plasma called the corona (\citealt{Eardley1975,Thorne1975}). 
This featureless radiation dominating the hard X-ray spectrum is often referred to as \textit{continuum} emission
and its shape can be successfully modelled with a phenomenological power law with an
exponential high energy cut-off correlated to the electron temperature of the plasma.
Some of the continuum photons travel \textit{directly} to the observer, while
some others illuminate the accretion disc, are reprocessed in the upper 
layers of its atmosphere and are re-emitted. 
This process is commonly called \textit{reflection} and 
its spectrum shows characteristic features such as the iron K$\alpha$ line 
at $6.4$ keV and the Compton hump peaking around $20-30$ keV 
(e.g. \citealt{Fabian1989}; \citealt{George1991}; \citealt{Garcia2010}).

The redshift due to the gravity of the black hole 
and the relativistic Doppler shifts due to the motion of the accretion disc broaden and 
skew the reflection features in the energy spectrum 
(e.g. \citealt{Fabian1989, Dauser2013}). 
These effects depend on characteristics of the system such as the black hole spin and the inner disc radius. 
Most of the current reflection models assume an electron density of
the accretion disc of $n_{\rm e} = 10^{15}$\,cm$^{-3}$ (e.g. \citealt{Ross2005, Garcia2010}). Even though this is arguably adequate for 
black holes with $M > 10^{7}\, M_{\odot}$ and high mass accretion rate, in
most other accreting black holes 
much higher densities are expected (even $>10^{20}$\,cm$^{-3}$ \citealt{Shakura1973}).
\citet{Garcia2016} noted that when the disc density is high 
the quasi-thermal component of the re-processed emission appears to peak in the soft X-ray band (increasing in effective temperature). 
Spectral fitting alone can lead to disagreements 
between different studies, in particular regarding measurement of the disc inner radius (e.g. compare \citealt{Parker2015} with \citealt{Basak2017} or 
\citealt{Garcia2015} with \citealt{Dzielak2019}).
This is due to the difficulties in disentangling the direct emission and
the reflected component (e.g. \citealt{Gierlinski1997, Basak2017}).

One of the most effective methods to study the X-ray emission in accreting 
black holes is spectral-timing analysis, combining the spectral 
information with the timing properties of the radiation. 
Timing analysis focuses on studying 
the variability of the light curve on relatively short timescales, generally of the order 
$\sim [10^{-3} - 10^{3}] M/M_{\odot}$ s (e.g. \citealt{McHardy2010}).
The Fourier transform (FT) allows us to study 
variability timescales in terms of Fourier frequencies assigning amplitudes and phases to each frequency.
The time variability arises mostly in the corona and it is transferred to the reflected emission. The different light travel times of 
the continuum, and the reflected radiation 
cause time lags that are measured as phase lags in a Fourier 
cross-spectrum analysis. 
Such 'reverberation' lags dominate the relatively high Fourier frequencies 
($\nu > 300 M_{\odot}/M$ Hz).
We expect the iron line and Compton hump energy bands, 
which are clearly dominated by  
reflection, to lag behind the rest of the spectrum.
These lags were first detected in Active Galactic Nuclei (AGN) 
(\citealt{McHardy2007,Fabian2009}), since the  
longer timescales allow us to detect many more photons 
per variability cycle compared to black hole binaries (BHBs). Since then the Fe line and Compton hump features have been observed in the lag-energy spectrum of several AGN (e.g. \citealt{Zoghbi2014,Kara2016}). 
In BHBs, reverberation lags were first detected through photons 
that are thermalised in the reprocessing mechanism and re-emitted 
as disc blackbody radiation (\citealt{Uttley2011,DeMarco2015}); 
only later the iron line feature was discovered in the lag-energy spectrum 
(\citealt{DeMarco2017, Kara2019}).

The transfer function formalism 
(e.g. \citealt{Campana1995,Reynolds1999,Wilkins2013}) 
encodes the relativistic dynamical distortions of the
reflected energy spectrum and the timing response 
of the reflected disc photons. 
Some of the reverberation models developed recently use this 
formalism to fit either the lag-frequency spectrum (e.g. \citealt{Emmanoulopoulos2014}) or the lag-energy spectrum (e.g. \citealt{Cackett2014}) without accounting for the rest-frame energy spectrum. 
However, a proper treatment of the flux-energy spectrum 
is needed to account for dilution of the reverberation lag, which occurs 
when the energy bands are not completely dominated by 
either the continuum or the reflected emission (\citealt{Uttley2014}).
This dilution effect is accounted for in the model \kynrefrev\ \citep{Dovciak2017,Caballero2018}, which 
uses the reflected rest-frame 
spectrum \relflionx\ (\citealt{Ross2005}) to calculate 
the lag-frequency spectrum. \citet{Alston2020} recently used this model to fit multi-epoch 
lag frequency spectra of IRAS~13224$-$3809 in order to break the degeneracy between 
the height of the corona and the mass of the black hole
that single-epoch fits to the lag-frequency spectrum suffer from.

All of these models focus on fitting the reverberation lags detected in AGN.  
However, these lags are not the only lags observed in accreting black holes: 
at lower Fourier frequencies ($\nu < 100 M_{\odot}/M$ Hz) these sources show the 
higher energy light curve variability lagging behind the 
lower energy light curve variability, a phenomenon often called `hard' lags.
These are characterised by a featureless lag-energy spectrum 
which has been detected in both BHBs and AGN 
(\citealt{Miyamoto1988, Nowak1999, Papadakis2001, McHardy2004}). 
These hard lags are thought to be caused by mass accretion 
rate fluctuations propagating through the 
accretion disc at the viscous timescale 
(e.g. \citealt{Lyubarskii1997, Kotov2001, Arevalo2006, Ingram2013}).

These two types of lags  contribute differently at each timescale.
Proper modelling them is crucial to constrain 
the parameters of the system.
\citet{Wilkins2016} proposed a model combining mass accretion rate fluctuations 
propagating either inwards or outwards in an extended corona with 
reverberation signal modelled through multiple transfer functions. 
Even though this is arguably one of the most complete models proposed to date, 
its large computational requirements make it impractical for use in an iterative fit procedure with current techniques.

The \reltrans\ model (\citealt{Ingram2019, Mastroserio2019} hereafter I19, M19) 
 accounts for the low and high frequency lags, modelling 
the systems through a point source corona (lamp-post) on the black hole spin axis and a razor-thin disc.
In order to account for the reverberation lags the model calculates
the different light travel times of the photons
with the relativistic ray-tracing code YNOGK (\citealt{Yang2013}). 
In the \reltrans\ model the continuum lags are accounted for by phenomenologically allowing the power-law 
spectral index and the power-law normalization to vary with time.
The idea is that the physical processes responsible for
heating and cooling the corona cause time variations of the power-law emission. 
Specifically, variations in the mass accretion rate from the disc to the corona and in the flux of 
disc seed photons hitting the corona cause the temperature and 
the optical depth of the coronal plasma to vary with time. Clearly, this description of the continuum spectral variations is phenomenological, and not derived from explicitly modeling these physical processes.
\reltrans\ parametrizes the time variations 
of both the power-law normalisation and its spectral index
using the analytical formalism  presented in \citet{Mastroserio2018} (hereafter M18).
The variations of the spectral index lead to a non-linear disc response that we account for self-consistently in the model by Taylor expanding the reflection component (M18).
\citet{Chainakun2016a} and \citet{Chainakun2017} proposed a `two-blob' 
geometry for the corona which leads to similar results. 

A common assumption of  reverberation models is that 
the disc responds linearly to the fluctuations of the illuminating flux. 
However, a variation in the illuminating flux has the effect 
of changing the ionization level of the disc atmosphere. 
In this work we include variations of the ionisation 
resulting from variations of the illuminating flux in our model. 
We show how these new feature changes the lag-energy spectrum and we test it on real data.
This new source of lags has never been considered in model fits to actual data. \citet{Mahmoud2019} discussed the variations of the 
ionisation state of the reflector and their effects on the lag-energy spectrum, but
only after the fitting procedure.  

In this paper we present the new model release \reltrans ~2.0.
In Section~\ref{sec:formalism} we describe the improved mathematical 
formalism showing the differences with the previous versions of the model.
In Section~\ref{sec:model_features} we investigate 
(i) the correlation among the parameters of the power-law time variation and 
their ranges given the observed variability,
(ii) how the reverberation lags change when we take into account 
high densities in the disc and (iii) the time variations of the ionisation of the disc due to the variations of the illuminating radiation.
As a proof of principle, in Section~\ref{sec:maxi1820} we fit the lag-energy spectrum of MAXI~J1820$+$070 with \reltrans ~2.0 focusing on 
how the different components contribute to the fit.
Finally, in Section~\ref{sec:discussion} we discuss why a pivoting power-law
approximation can describe the expected spectral time variations and how 
considering reflection from a higher density accretion disc together with 
ionisation time-variations contribute to fit the observed low energy lags. 
We conclude with some remarks on the lag energy spectral fitting.

\section{ reltrans~Version 2.0}
\label{sec:formalism}

In this Section we describe the \reltrans~version 2.0 formalism. We first summarise version 1.0, which is the previous publicly available version\footnote{https://adingram.bitbucket.io/reltrans.html} described in I19. We then summarise the improvements that have been made from version 1.0 to 2.0, before describing the mathematical details of the continuum and reflection components in the new version.

\subsection{Basic equations}
\label{sec:v1}

\reltrans~v1.0 assumes that the corona is a spherical, isotropic, point-like source (i.e. a lamp-post). The surface area of the source is $a_s$ and the time-dependent specific flux it emits in its own restframe is
\begin{equation}
    F_s(E_s,t) = \frac{C(t)}{a_s} E_s^{1-\Gamma} {\rm e}^{-E_s/E_{\rm cut}},
    \label{eq:lamp-post_specific_flux}
\end{equation}
where $C(t)$ is the normalisation that varies with time, $\Gamma$ is the photon index,  $E_s$ and $E_{\rm cut}$ are respectively the photon energy and high energy cut-off, both in the source rest frame. An alternate flavour of the model, \reltransCp, uses the thermal Comptonisation model \nthcomp~\citep{Zdziarski1996} instead of an exponentially cut-off power-law. The FT of the specific flux seen by the observer as a function of energy in the observer's rest frame is then 
\begin{equation}
    F(E,\nu) = A(\nu) \ell g_{\rm so}^{\Gamma} E^{1-\Gamma} {\rm e}^{-E/E_{\rm cut,obs}},
    \label{eq:continuum_first}
\end{equation}
where $A \equiv C/(4\pi D^2)$, $D$ is distance from the observer to the source, $g_{\rm so}=E/E_s$ is the blueshift experienced by photons as they travel from the source to the observer and $\ell$ accounts for gravitational lensing\footnote{Note that in I19, cosmological redshift was included explicitly whereas here we absorb it into the definitions of $g_{\rm so}$, $\tau_{\rm so}$, $g_{\rm do}$ and $\tau_{\rm do}$.}.

The FT of the direct observed specific flux of the reflection component is $R(E,\nu) = (1/\mathcal{B}) A(\nu) W(E,\nu)$, where $W(E,\nu)$ is the transfer function and the ratio $1/\mathcal{B}$ is the `boost' parameter, which enables the model to account for deviations from our assumptions of an isotropically radiating source and point-like geometry. The transfer function is given by
\begin{equation}
    W(E,\nu) = \int_{\alpha,\beta} g_{\rm do}^3(r,\phi) \epsilon(r) {\rm e}^{i 2\pi \tau(r,\phi) \nu} \mathcal{R}[ E / g_{\rm do}(r,\phi) ] {\rm d} \alpha {\rm d} \beta.
    \label{eqn:Wonezone}
\end{equation}
Here $\alpha$ and $\beta$ (the impact parameters at infinity) are the projected horizontal and vertical distances on the image plane from the black hole to a point on the disc with radius $r$ and azimuthal angle $\phi$. $g_{\rm do}(r,\phi)$ is the energy shift experienced by photons travelling from disc coordinate $r,\phi$ to the observer. $\epsilon(r)$ is the radial emissivity profile (Equation 20 in I19). A grid of impact parameters are converted to $r$ and $\phi$ coordinates by ray-tracing backwards in time in the Kerr metric (\citealt{Dexter2009,Yang2013}) and the integral is for all impact parameter combinations for which $r \in [r_{\rm in},r_{\rm out}]$. The reverberation lag, $\tau(r,\phi)$, is the extra time taken by photons that reflect from the disc at coordinates $r,\phi$ before reaching the observer, with respect to the photons that travelled directly from the corona to the observer.
The emergent reflection spectrum (specific intensity as a function of energy) $\mathcal{R}(E)$ is calculated using \xillver~(or \xillverCp~in the case of \reltransCp; \citealt{Garcia2013}).

Direct calculation of eq.~(\ref{eqn:Wonezone}) is computationally prohibitive because $\mathcal{R}(E)$, in general, depends on disc coordinates $\alpha$ and $\beta$. The simplest solution is to ignore all dependencies of $\mathcal{R}(E)$ on disc coordinates and consider a single rest-frame reflection spectrum for the entire accretion disc.  
This `one zone' model has several limitations: i) The emergent reflection spectrum, $\mathcal{R}(E)$, depends on the initial trajectory of the photons as they emerge from the disc -- the emission angle $\theta_e$; $\mu_e \equiv \cos\theta_e$ -- which is different for different disc coordinates due to light bending \citep{Garcia2014}. ii) The energy shift $g_{\rm sd}(r)$ experienced by photons travelling from source to disc means that different disc radii see a different illuminating spectrum. iii) The ionization parameter $\xi(r) = 4 \pi F_x(r)/n_{\rm e}(r)$, can depend on disc radius. A more correct treatment is to divide the accretion disc in multiple zones. The transfer function is therefore more accurately given by
\begin{equation}
    W(E,\nu) = \sum_j^J \sum_{k}^K \Delta W(E,\nu|\mu_e(j),g_{\rm sd}(r_k),\log_{10}\xi(r_k)),
\end{equation}
where we define $J$ bins for $\mu_e$, $K$ radial bins, and $\Delta W(E,t|j,k)$ is given by eq.~(\ref{eqn:Wonezone}) except that the integral is only over impact parameter combinations for which $\mu_e$ is in the $j^{\rm th}$ bin and $r$ is in the $k^{\rm th}$ bin.
$J$ and $K$ are set by the environment variables \texttt{MU\_ZONES} and \texttt{ION\_ZONES}\footnote{Note that v1.0 also included the environment variable \texttt{ECUT\_ZONES} that determined the number of $g_{\rm sd}$ zones independently of the number of $\xi$ zones. This environment variable is now obsolete in v2.0.}. 

The FT of the total spectrum (direct plus reflected) is then simply $S(E,\nu) = F(E,\nu)+R(E,\nu)$, and the cross-spectrum is $G(E,\nu) = S(E,\nu) F_r^*(\nu)$, where $F_r^*(\nu)$ is the FT of the count rate in the reference band. This is the count rate summed over a range of energy channels, and so $F_r(\nu)$ can be calculated from $S(E,\nu)$ and the response matrix of the X-ray detector (see I19 and \citealt{Mastroserio2020} for more details).

\subsection{Summary of improvements}

Version 2.0 includes the improvements introduced in M19 and \cite{Mastroserio2020} but not hitherto included in a public release:

\noindent \textbf{Continuum lags:} We follow M19 by including fluctuations in the photon index, $\Gamma$, using the Taylor expansion formalism of \cite{Mastroserio2018}. These fluctuations give rise to continuum lags.

\noindent \textbf{Frequency averaging:} We follow \citet{Mastroserio2020} by first calculating the cross-spectrum as a function of Fourier frequency and only then averaging over the input frequency range. Version 1.0 instead first averaged the transfer function over the frequency range and then calculated the cross-spectrum from that (see the discussion in \citealt{Mastroserio2020} for more details).

\paragraph*{Pivoting parameters:} In M18 and M19, the fluctuations in $\Gamma$ that cause continuum lags in the model are parameterised by the parameters $\phi_B(\nu)$ and $\gamma(\nu)$ (they are going to be defined in Section~\ref{sec:formalism_continuum}). In \citet{Mastroserio2020} we slightly changed the definitions of these continuum lag parameters to link them to the continuum variability in a much more intuitive way. Here we investigate these pivoting parameters and their impact on the lag-energy spectrum.
\\

\noindent We also introduce several features here for the first time:
\paragraph*{High density disc:}
The \xillver~and \xillverCp~grids assume an electron number density of $10^{15}$\,cm$^{-3}$. This is appropriate for an AGN with a $M > 10^7~M_\odot$ black hole, whereas the disc density is expected to be orders of magnitude higher in X-ray binary discs. For version 2.0, we therefore use the publicly available \xillverD\ and \xillverDCp\ tables \citep{Garcia2016} for the restframe reflection spectrum. The disc density is now a model parameter that can be varied in the range $n_{\rm e}=10^{15-20}\,{\rm cm}^{-3}$.

\paragraph*{Ionization parameter fluctuations:} 
An increase in illuminating flux will increase the ionization parameter and thus adjust the ionization balance in the disc atmosphere, changing the shape of the emergent reflection spectrum. Here, we account for the variations in $\xi(t)$ that result from variations in the source spectrum by using the first order Taylor expansion formalism already used to account for $\Gamma$ variations.

\paragraph*{Performance:} The new version uses a more efficient fast FT routine (\citealt{Frigo2005}) for the convolutions that are computed in the calculation of the transfer functions. The code is now $5$ times faster than the 
previous versions.
\\

Finally, we note that, even though the model accounts for additional physics, the only new parameter is the disc density. 

\subsection{Continuum}
\label{sec:formalism_continuum}
Here we present the mathematical formalism of the continuum emission in \reltrans .
For a time-dependent photon index, the observed continuum spectrum becomes
\begin{equation}
    F(E,t) = A(t) \ell g_{\rm so}^{\Gamma(t)} E^{1-\Gamma(t)} {\rm e}^{-E/E_{\rm cut,obs}},
    \label{eq:continuum_first}
\end{equation}
which is identical to the expression of I19 except for the time-dependence of $\Gamma$ (a similar expression can be written down for the \nthcomp~flavour of the model). Taylor expanding eq.~\ref{eq:continuum_first} around $\Gamma=\Gamma_0$ and keeping only first order terms gives
\begin{equation}
    F(E,t) \approx F(E,t)\bigg|_{\Gamma=\Gamma_0} + \frac{\partial F(E,t)}{\partial \Gamma}\bigg|_{\Gamma=\Gamma_0} [\Gamma(t)-\Gamma_0].
\end{equation}
Defining $D(E) \equiv \ell g_{\rm so}^{\Gamma_0} E^{1-\Gamma_0} {\rm e}^{-E/E_{\rm cut,obs}}$, we can simplify eq.~\ref{eq:continuum_first} by writing $F(E,t)=A(t)D(E)g_{\rm so}^{\delta\Gamma(t)} E^{-\delta\Gamma(t)}$, where $\delta\Gamma(t)=\Gamma(t)-\Gamma_0$. This leads to
\begin{equation}
    F(E,t) \approx A(t) D(E) \left[ 1 - \ln(E/g_{\rm so}) \delta \Gamma(t)\right].
    \label{eq:continuum_taylor_exp}
\end{equation}
Defining $A(t) = A_0 + \delta A(t)$, eq.~\ref{eq:continuum_taylor_exp} becomes
\begin{equation}
    F(E,t) \approx A_0 D(E) \left[ 1 + \delta A(t)/A_0 - \ln(E/g_{\rm so}) \delta \Gamma(t)\right],
    \label{eqn:F}
\end{equation}
where we have ignored the term of order $\delta A \delta\Gamma$.
The FT of this, for Fourier frequency $\nu>0$, is
\begin{equation}
    F(E,\nu) \approx D(E) \left[ A(\nu) - A_0 \ln(E/g_{\rm so}) \Gamma(\nu)\right].
    \label{eqn:Fnu}
\end{equation}
Following \citet{Mastroserio2020}, we define $\gamma(\nu) {\rm e}^{i \phi_{AB}(\nu)} \equiv A_0 \Gamma(\nu) / A(\nu)$ to give
\begin{equation}
    F(E,\nu) \approx A(\nu) D(E) \left[ 1 - \ln(E/g_{\rm so}) \gamma(\nu) {\rm e}^{i \phi_{AB}(\nu)} \right].
    \label{eqn:Fnu2}
\end{equation}
Here, $\phi_{AB}(\nu)$
is the phase difference between fluctuations in $\Gamma(t)$ and those in $A(t)$ (positive means that $\Gamma$ lags $A$). We specify $\phi_{AB}(\nu)$ (averaged over a frequency range: see \citealt{Mastroserio2020}) as a model parameter
\footnote{Previous versions of the model used a parameter referred to in M18 and M19 as $\phi_B(\nu)$. The two parameters relate to each other as $\phi_{AB}(\nu)=\phi_{B}(\nu)+\pi-\phi_A(\nu)$.}.
The parameter $\gamma(\nu) = A_0 |\Gamma(\nu)|/|A(\nu)|$ has the same meaning as in M19, except we now explicitly ignore in its definition terms of the order $\delta A \delta\Gamma$ and higher.

\subsection{Reflection}
\label{sec:formalism_reflection}

The reflection spectrum from a patch of disc that subtends a
solid angle $d\alpha d\beta/D^2$ is (for unity boost 
parameter, i.e., an isotropic continuum source)
\begin{equation}
    dR(E,t) = A(t') g_{\rm do}^3(r,\phi) \epsilon(r,t') \mathcal{R}[E/g_{\rm do}|\xi(t'),\Gamma(t')] d\alpha d\beta,
    \label{eq:taylor_exp_reflection}
\end{equation}
where $t' = t - \tau(r,\phi)$. Hereafter, explicit $r$ and $\phi$ dependencies will be omitted for brevity. As $A(t)$ and $\Gamma(t)$ fluctuate in time, $dR(E,t)$ will also fluctuate. Our previous treatment did not account for the changes in the ionization parameter $\xi(t)$ that result from fluctuations 
in the primary continuum expressed by $A(t)$ and in $\Gamma(t)$.
Here, we will also account for these changes. Taylor expanding $dR$ around $A=A_0$ and $\Gamma=\Gamma_0$ and only keeping linear terms gives
\begin{equation}
\begin{split}
    dR(E,t) \approx dR\bigg|_{A=A_0,\Gamma=\Gamma_0} + \frac{\partial (dR)}{\partial A}\bigg|_{A=A_0,\Gamma=\Gamma_0} \delta A(t') \\
    + \frac{\partial (dR)}{\partial \Gamma}\bigg|_{A=A_0,\Gamma=\Gamma_0} \delta \Gamma(t'),
\end{split}
    \label{eq:refl_Taylor}
\end{equation}
where $\delta\Gamma(t)=\Gamma(t)-\Gamma_0$. Evaluating the first differential gives
\begin{equation}
    \frac{\partial (dR)}{\partial A} = g_{\rm do}^3 \epsilon d\alpha d\beta \left[ \mathcal{R}(E_d) + A \frac{\partial \mathcal{R}(E_d)}{\partial A} \right].
    \label{eq:refl_norm_expansion1}
\end{equation}

The \textsc{xillver} model grids are not tabulated in terms of $A$. The model parameter is instead the logarithm of the ionization parameter $\xi$, which is
expressed as the ratio between the incident flux and the disc
density ($\xi = 4\pi F_x / n_{\rm e}$). 
In order to numerically evaluate the differential $\partial \mathcal{R} / \partial A$ in the above equation, we therefore must re-express $\partial/\partial A$ in terms of $\partial / \partial \log\xi$, where `$\log$' is taken to mean logarithm to the base $10$ throughout. It is straight forward to show from the chain rule that
\begin{equation}
    \frac{\partial}{\partial A} = \frac{\partial \xi}{\partial A} \frac{\partial}{\partial \xi} = \frac{\partial \xi}{\partial A}   \frac{\partial \log\xi}{\partial \xi} \frac{\partial}{\partial \log\xi} = \frac{\partial \xi}{\partial A} \frac{1}{\ln 10} \frac{1}{\xi}  \frac{\partial}{\partial \log\xi}.
    \label{eqn:chainrule}
\end{equation}
We therefore must evaluate $\partial\xi / \partial A$.
We assume the density is constant over time, thus the ionisation parameter is given by
\begin{equation}
\xi(t) = \frac{A(t)}{A_0} ~\xi_0 ~g_{\rm so}^{\delta\Gamma(t)} S(t),
\end{equation}
where $S(t)$ is the time variation of the incident flux between 
$0.1$ keV and $1$ MeV, which is written as
\begin{equation}
    S(t) = \frac{ \int_{ 0.1 {\rm keV} }^{ 1 {\rm MeV} } E^{-\delta\Gamma(t)} E^{1-\Gamma_0} {\rm e}^{-E/E_{\rm cut,o}} dE } {\int_{ 0.1 {\rm keV} }^{ 1 {\rm MeV} } E^{1-\Gamma_0} {\rm e}^{-E/E_{\rm cut,o}} dE},
    \label{eq:ionisation_variation}
\end{equation}
and $\xi_0$ is the 
time-averaged ionization parameter. 
In eq.~\ref{eq:ionisation_variation} the continuum is represented by a cut off power-law, however, in the model where the continuum is \nthcomp , that spectrum is substituted for the cut-off power law.
Eq. \ref{eqn:chainrule} therefore becomes
\begin{equation}
    \frac{\partial}{\partial A} = \frac{1}{\ln 10~A} \frac{\partial}{\partial \log\xi}.
\end{equation}
We can therefore apply the above to eq. \ref{eq:refl_norm_expansion1} to obtain
\begin{equation}
        \frac{\partial (dR)}{\partial A} = g_{\rm do}^3 \epsilon d\alpha d\beta \left[ \mathcal{R}(E_d) + \frac{1}{\ln 10} \frac{\partial \mathcal{R}(E_d)}{\partial \log\xi} \right],
        \label{eq:refl_norm_expansion2}
\end{equation}
and the partial derivative $\partial \mathcal{R}/\partial \log\xi$ is simple to evaluate numerically from the relevant \textsc{xillver} model.

Now we evaluate the derivative of $dR$ with respect to $\Gamma$. Following M19, we can write
\begin{equation}
    \frac{\partial (dR)}{\partial \Gamma}\bigg|_{A=A_0,\Gamma=\Gamma_0} = A_0 g_{\rm do}^3 \epsilon d\alpha d\beta \left[ \ln g_{\rm sd} \mathcal{R}(E_d) + \frac{\partial \mathcal{R}(E_d)}{\partial \Gamma}  \right].
    \label{eq:refl_gamma_expansion1}
\end{equation}
As a slight subtlety, we note that $\partial \mathcal{R}/\partial \Gamma$ is evaluated as
\begin{equation}
    \frac{\partial \mathcal{R}(E)}{\partial \Gamma} \approx \frac{ \mathcal{R}(E|\Gamma_2,\log\xi_2)-\mathcal{R}(E|\Gamma_1,\log\xi_1) }{ \Gamma_2 - \Gamma_1 },
    \label{eq:refl_gamma_expantion2}
\end{equation}
where $\Gamma_2=\Gamma_0+\Delta\Gamma/2$, $\Gamma_1=\Gamma_0-\Delta\Gamma/2$, $\log\xi_2 = \log\xi_0 + \log g_{\rm so}~\Delta \Gamma/2 + \log S_2$, $\log\xi_1 = \log\xi_0 - \log g_{\rm so}~\Delta \Gamma/2 + \log S_1$, $S_2=S(t)\big|_{\delta\Gamma=\Delta\Gamma/2}$ and $S_1=S(t)\big|_{\delta\Gamma=-\Delta\Gamma/2}$. Here, $\Delta\Gamma$ is an internal model resolution parameter.

We combine the equations above (\ref{eq:refl_Taylor}-\ref{eq:refl_gamma_expansion1}) and explicitly include the boost parameter to obtain a linearised expression for $dR$,
\begin{equation}
\begin{split}
    dR(E,t) \approx \frac{1}{\mathcal{B}}~g_{\rm do}^3 \epsilon d\alpha d\beta \bigg\{ 
    A(t') \mathcal{R}(E_d) + \frac{\delta A(t')}{\ln 10} \frac{\partial\mathcal{R}(E_d)}{\partial \log\xi} \\
    + A_0 \delta\Gamma(t') \log g_{sd} \mathcal{R}(E_d) + A_0 \delta\Gamma(t') \frac{\partial\mathcal{R}(E_d)}{\partial \Gamma}
    \bigg\}
    \end{split}
\end{equation}
Integrating over all disc patches gives the total observed reflection spectrum. The FT of the reflection spectrum, for Fourier frequency $\nu>0$, is
\begin{equation}
\begin{split}
    R(E,\nu) \approx \, & (1/\mathcal{B}) \left\{ ~A(\nu) [ W_0(E,\nu) + W_3(E,\nu) ] \right.\\
    &  \left.+ A_0 \Gamma(\nu) [ W_1(E,\nu) + W_2(E,\nu)] \right\},
    \label{eqn:R1}
    \end{split}
\end{equation}
where the transfer functions (in the `one zone' limit, see eq.~\ref{eqn:Wonezone}) are
\begin{eqnarray}
W_0(E,\nu) &=& \int_{\alpha,\beta} g_{\rm do}^3 \epsilon {\rm e}^{i 2\pi \tau \nu} \mathcal{R}(E_d) d\alpha d\beta \\
W_1(E,\nu) &=& \int_{\alpha,\beta} g_{\rm do}^3 \epsilon {\rm e}^{i 2\pi \tau \nu} \ln g_{\rm sd} \mathcal{R}(E_d) d\alpha d\beta \\
W_2(E,\nu) &=& \int_{\alpha,\beta} g_{\rm do}^3 \epsilon {\rm e}^{i 2\pi \tau \nu} \frac{\partial \mathcal{R}}{\partial \Gamma}(E_d) d\alpha d\beta \\
W_3(E,\nu) &=& \frac{1}{\ln 10} \int_{\alpha,\beta} g_{\rm do}^3 \epsilon {\rm e}^{i 2\pi \tau \nu} \frac{\partial \mathcal{R}}{\partial \log\xi}(E_d) d\alpha d\beta.
\end{eqnarray}
$W_3$ accounts for the ionization parameter fluctuations, whereas $W_1$ and $W_2$ account for the photon index variations. 
Note that transfer function $W_0$ is identical to the transfer function in I19 (our Equation \ref{eqn:Wonezone}), transfer functions $W_0$, $W_1$ and $W_2$ are defined in exactly the same way as in M19, and only transfer function $W_3$ is new. For the multi-zone limit, in addition to $\mu_e$ depending on $r$ and $\phi$ and $g_{\rm sd}$ and $\xi$ depending on radius, we can now also account for $n_{\rm e}$ depending on radius since it is an input parameter of the \xillverD~model. The full expression for each transfer function becomes
\begin{equation}
    W(E,\nu) = \sum_j^J \sum_k^K \Delta W(E,\nu|\mu_e(j),g_{\rm sd}(r_k),\log_{10}\xi(r_k),n_{\rm e}(r_k)),
\end{equation}
where we define $J$ bins for $\mu_e$ and $K$ radial bins, and $\Delta W(E,t|j,k)$ is defined analogously to its equivalent in Section \ref{sec:v1}. We can re-write eq.~\ref{eqn:R1} in terms of $\gamma(\nu)$ and $\phi_{AB}(\nu)$ to give
\begin{equation}
\begin{split}
    R(E,\nu) \approx \, & (1/\mathcal{B})A(\nu) \left\{ W_0(E,\nu) + W_3(E,\nu) \right.\\
    &  \left.+ \gamma(\nu) {\rm e}^{i \phi_{AB}(\nu)} [ W_1(E,\nu) + W_2(E,\nu)] \right\}.
    \label{eqn:R2}
    \end{split}
\end{equation}

\subsection{Total cross-spectrum}

As described above, the cross-spectrum is calculated as $G(E,\nu)=S(E,\nu) F_r^*(\nu)$, where $S(E,\nu) = F(E,\nu) + R(E,\nu)$ is the total spectrum and $F_r(\nu)$ is the FT of the reference band count rate, which we calculate self-consistently from our model for $S(E,\nu)$.
This process constrains the phase of $A(\nu)$, since the phase lag between the reference band and itself must be zero. We then average the cross-spectrum over the frequency range specified by the user in the manner described in \cite{Mastroserio2020}. The amplitude of $A(\nu)$ is specified by the model parameter $\alpha(\nu)$, which is a weighted average of $|A(\nu)|$ across the specified frequency range. Finally, we multiply the frequency-averaged cross-spectrum by ${\rm e}^{i \phi_A(\nu)}$, which accounts for imperfect calibration. If the response matrix perfectly describes the true response of the detector, then $\phi_A$ will always be zero. However, we may wish to include poorly calibrated energy channels in the reference band in order to increase signal to noise, in which case it may be necessary to leave $\phi_A$ as a free parameter (although it should always be small). 

The free continuum parameters in the model, for each frequency range, are $\alpha$ (which is the \xspec\ \texttt{norm} parameter), $\phi_{AB}$, $\gamma$, and optionally $\phi_{A}$. 
For the DC ($\nu=0$) component, we instead have the much simpler expression $G(E) = \alpha_0 \left[ D(E) + (1/\mathcal{B}) W_0(E)  \right]$. It is worth probing the effect of the new transfer function ($W_3$)
on the lag spectrum and using the slightly changed continuum formalism 
to investigate the impact of the continuum variations. Section~\ref{sec:model_features} is dedicated to this task. 

\subsection{Linear approximation}
We use a first order Taylor expansion to approximate the changes in the reflection spectrum as a function of power-law index and ionisation variations. This approximation holds when either the variations are small, or the reflected spectrum depends linearly on these two parameters. Usually in accreting stellar-mass black holes the combined variability of power-law photon index and the normalisation of the incident continuum causes flux variations that are less than $40\%$ of the total flux on the timescales where the reverberation lags dominate ($>1$ Hz). However, some AGN show strong (factors of a few) variability on the reverberation time scales (see e.g. 1H~0707–495: \citealt{Kara2013a, Pawar2017} and IRAS~13224–3809: \citealt{Alston2019}). 
The remainder of the Taylor expansion depends on how non-linear the reflection response is, which depends on photon energy. For energies where electron scattering is the dominant source of opacity, such as above 10\,keV or so,  where the Compton hump dominates, the reflected emission responds linearly to the variation of the incident flux. At energies where true absorption and line emission are dominant, which are important for constraining the reverberation parameters, however, the response is non-linear.
Thus, in particular for those AGN where the incident continuum varies more than $100\%$ on the reverberation timescales, we suggest we suggest to use the model with caution, since the first order Taylor expansion may not be a good approximation.

In the following we give a more mathematical treatment of the limitations of the linear approximation.
In eq.~\ref{eq:taylor_exp_reflection} there are quadratic terms of the Taylor expansions that were ignored, such as
\begin{equation}
    q(t_k) = \frac{\partial^2R}{\partial A^2}~\delta A^2(t_k)
\end{equation}
where we have expressed $A(t)$ as a discrete time series instead of a continuus function. The second differential $\partial^2 R/\partial A^2$ is zero for electron scattering, but non-zero for absorption and fluorescence. We wish to determine the contribution of $q(t_k)$ to the cross-spectrum, $\langle Q(\nu_j) F_r^*(\nu_j) \rangle$, where the angle brakets denote averaging over $L$ realizations. From the convolution theorem, we can write
\begin{equation}
    Q(\nu_j) \propto \sum_{k=-N/2+1}^{N/2} A(\nu_{j-k}) A(\nu_k),
\end{equation}
and thus
\begin{equation}
    \langle Q(\nu_j) F_r^*(\nu_j) \rangle \propto \sum_{k=-N/2+1}^{N/2} \frac{1}{L} \sum_{\ell=1}^L A(\nu_{j-k}) A(\nu_k) F_r^*(\nu_j).
\end{equation}
For illustration, we can approximate $F_r(\nu_j) \propto A(\nu_j)$ to give
\begin{eqnarray}
    \langle Q(\nu_j) F_r^*(\nu_j) \rangle &\propto& \sum_{k=-N/2+1}^{N/2} \frac{1}{L} \sum_{\ell=1}^L F_r(\nu_{j-k}) F_r(\nu_k) F_r^*(\nu_j) \nonumber \\
    & \propto & \sum_{k=-N/2+1}^{N/2} B(\nu_{j-k},\nu_k),
\end{eqnarray}
where
\begin{equation}
    B(\nu_j,\nu_k) = \frac{1}{L} \sum_{\ell=1}^L F_r(\nu_{j}) F_r(\nu_k) F_r^*(\nu_{j+k}),
\end{equation}
is the \textit{bi-spectrum} of the reference band flux \citep[e.g.][]{Maccarone2013}. The importance of the quadratic term therefore depends on the \textit{bi-coherence} of the X-ray signal, which is the magnitude of the bi-spectrum \citep{Maccarone2011}. The above gives us an intuitive way of understanding the bi-coherence: if a time series has zero bi-coherence, the power spectrum of the square of the time series will be zero. For AGN and X-ray binaries, the bi-coherence is non-zero but it is typically $\lesssim 1\%$ \citep[e.g.][]{Uttley2005,Maccarone2011}. Therefore the contribution of the quadratic terms in the Taylor expansion is likely small \textit{even} in cases whereby the second differential function is large.

\section{Model features}
\label{sec:model_features}
In this section we visualise the characteristics of the \reltrans~2.0 model
in connection to the changes in the mathematical formalism, 
and how it compares to the previous version. In all the tests shown here, we use the 
default  parameters listed in Table~\ref{tab:defaut_param}, unless stated otherwise.
We first focus on the continuum lags that dominate at relatively low Fourier frequencies. We study how the continuum parameters 
$\phi_{AB}(\nu)$ and $\gamma(\nu)$, respectively the phase
difference and the amplitude ratio between $\Gamma(t)$ 
and $A(t)$, affect the predicted lag-energy spectrum. 
We then analyse the reverberation lags that dominate at relatively 
high Fourier frequencies. 
We study the effect of high electron density, up to $n_{\rm e} = 10^{20}$\,cm$^{-3}$,
in the accretion disc on the lag-energy spectrum, comparing to the 
old version that had this parameter fixed to $10^{15}$\,cm$^{-3}$. 
At the same time, we study how the time fluctuations of the ionisation 
parameter, 
that are encoded in the $W_3$ transfer function, 
impact on the calculation of the lag, and how these non-linear effects
combine with the radial dependence of the ionisation parameter, 
assuming either constant radial profile or a self-consistently calculated one.

\subsection{Continuum variability}
\label{sec:pivoting_parameters}

\begin{figure*}
	\includegraphics[width=\textwidth]{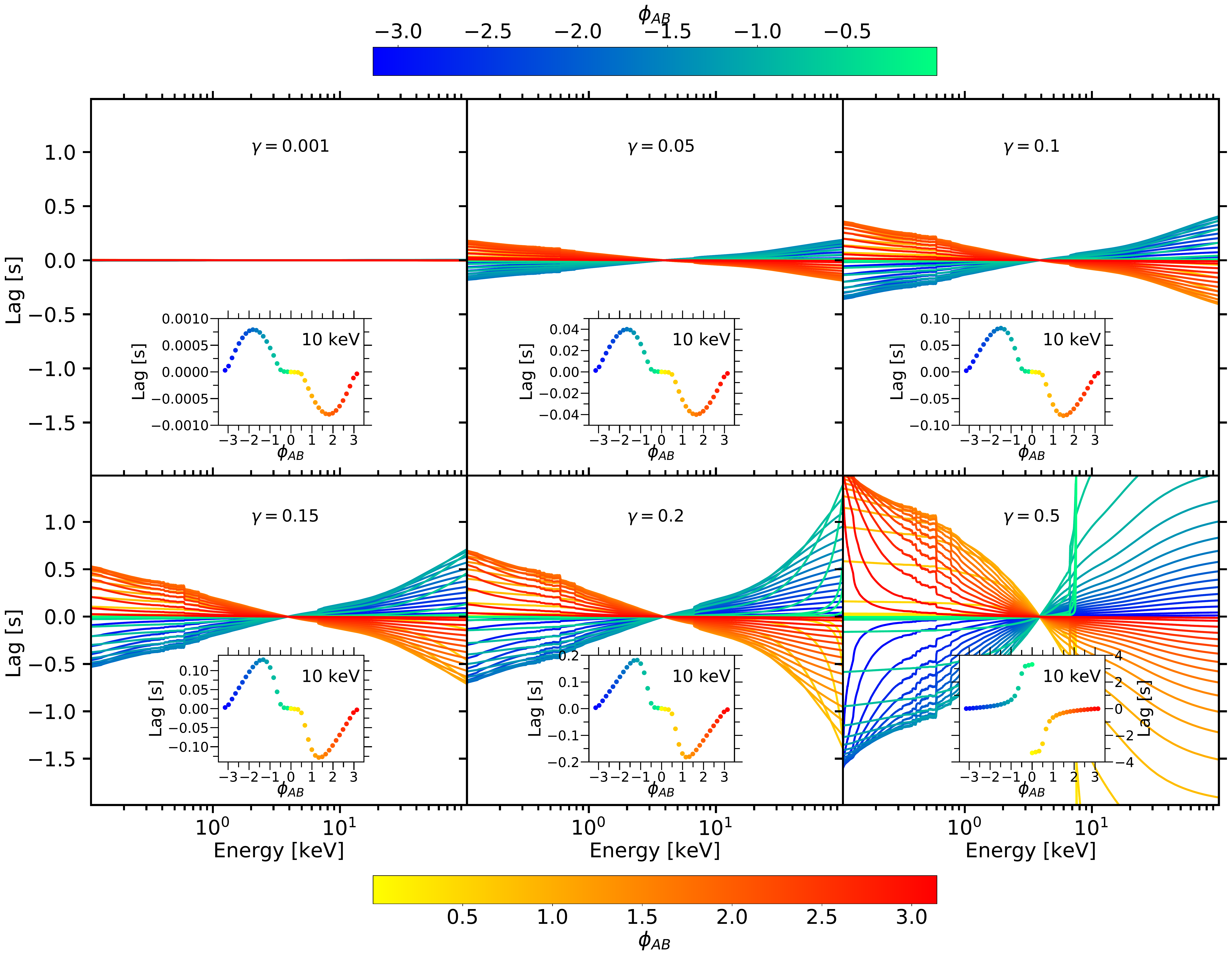}
    \caption{Main panels: lag energy spectra computed by the \reltransDCp\  
    model in the frequency range $0.1-0.2$ Hz varying the continuum parameters $\gamma$ and $\phi_{AB}$. They are respectively the amplitude ratio and the phase difference between the power-law index variability ($\Gamma(t)$) and power-law normalization variability ($A(t)$) . 
    The value of $\gamma$ is specified for each panel and $\phi_{AB}$ is color coded 
    following the color bars above and below the panels, the parameter ranges between 
    $-\pi$ and $\pi$. 
    The rest of the parameters have default values 
    (see Table~\ref{tab:defaut_param}). 
    The insets show the lag at $10$ keV as a function of $\phi_{AB}$ while $\gamma$ 
    is the same of the main panel. Each point in the inset plots correspond to the 
    value of each lines in the corresponding panels, they have the same color code. 
    }
    \label{fig:pivoting_lags}
\end{figure*}

We consider the Fourier frequency range $0.1$ - $0.2$ Hz
for a $10 M_{\odot}$ black hole. 
In this frequency range, the observed lags are dominated by the continuum lags, 
and the reverberation contribution is rather small. 
Therefore the lag-energy spectrum is mainly affected by the values of 
$\phi_{AB}$ and $\gamma$.
The former is the phase difference between the time variations of the power-law index $\Gamma$
and the normalisation $A$. 
Negative values mean that $A$ lags $\Gamma$, hence the peak of the hardness follows 
the peak in the normalisation, since higher $\Gamma$ corresponds to a softer spectrum and $A$ is proportional to the specific flux at 1 keV. 
This mechanism produces hard lags 
in the lag-energy spectrum. 
Conversely, positive values of $\phi_{AB}$ mean that 
$\Gamma$ lags $A$, resulting in soft lags.
In general, $\phi_{AB}$ values can range between $-\pi$ and $\pi$, 
since we do not have any physically motivated reason to further constrain 
this range.

The $\gamma$ parameter, instead, is related to the amplitude of the 
continuum flux variation. 
More specifically, following the definition at the end of
Section~\ref{sec:formalism_continuum} we can write 
$|\Gamma(\nu)| = \gamma(\nu) |A(\nu)|/ A_0$.  
This expression allows us to determine physically plausible constraints
of the $\gamma$ parameter. 
We note that relatively large values of $\gamma$ would break our fundamental 
assumption of being able to linearise 
(see the Appendix in M18), 
and they would predict extreme variability 
in some X-ray band because of the large pivoting variation.
This energy dependent extreme variability is not observed in the data.

Since a typical black hole binary in the hard state 
shows fractional root mean square (\textit{rms}) variability in its X-ray flux up to 
$20-40\%$, and the main source of 
the continuum variability is due to the variation of 
the normalisation, we can assume $|A(\nu)|/ A_0$ 
is at least $10\%$ during the hard to intermediate state. 
In the case of an AGN the variability is usually lower, 
typically $\sim 10\%$, even though the value can differ 
much depending on the system. 
Accreting black holes usually display $\Gamma$ ranging between $1.4$ and $2.6$, 
therefore $\gamma(\nu) = 0.5$ corresponds to a fractional variability 
amplitude in $\Gamma$ of $\sim 2.5\%$, which is reasonable if we assume 
that part of the continuum variability is also due to 
power-law index variations.
We therefore consider $\gamma$ ranging between $~0$ and $0.5$, 
since the upper limit is still small enough to not violate our 
assumptions or obviously contradict observations.
However, these limits are not strictly confirmed by the observations. 
The most accurate method to calculate variability
of $\Gamma$ is to fit a time series of flux energy spectra 
with the time resolution equal to the timescale considered in the lag analysis
(Steiner et al. in prep.). 

Fig.~\ref{fig:pivoting_lags} shows the lag as a function of energy computed for 
different values of $\gamma$ and $\phi_{AB}$ with the \reltransDCp\ model 
in the Fourier frequency range of $0.1$ - $0.2$ Hz for a $10 M_{\odot}$ black hole. 
The reference band is the entire energy band $0.1$ - $100$ keV, hence there must be one portion of the spectrum 
that varies as the reference band, and at that point the phase lag is zero by definition.
However, this is not the pivoting point of the continuum emission which does not necessarily exist because we account for the variations in both the power-law index and normalisation.  
We see from Fig.~\ref{fig:pivoting_lags} that increasing 
$\gamma$ results in longer lags.
This is not surprising, since increasing $\gamma$ enlarges 
the power-law index range and enhances its variability. 
When $\phi_{AB} < 0$ the model simulations show hard lags (i.e. lag increases with energy) and vice-versa when $\phi_{AB} > 0$, as expected. 
The inset in each panel shows the value of the lag of each line at $10$ 
keV as a function of $\phi_{AB}$. 
We note that the same value of the lag occurs twice in each inset, for two separate values of $\phi_{AB}$. 
This degeneracy is broken by considering 
the shape of the overall lag-energy spectrum.
In the inset of the bottom right panel ($\gamma = 0.5$), the lag reaches $\sim \pm 3.3$\,s at $\phi_{AB} = 0$, which corresponds to a phase difference of $\pm \pi$ (which are indistinguishable from each other) at the considered Fourier frequency of $0.15$ Hz. This means that the $10$ keV flux is varying in anti-phase with the reference band flux, which is the opposite of what happens in all the other insets. This is because at high values of $\gamma$ (such that $\gamma \log(E/g_{so}) > 1 $) the spectral index variations dominate over the normalisation variations and the we see anti-phase variations between high energies and low energies for the $\phi_{AB}=0$ case. 
When $\gamma$ is small, ($\gamma \log(E/g_{so}) < 1 $), the normalisation variations dominate and we see zero lag between each energy band and the reference band in the $\phi_{AB}=0$ case.

We note that, even though continuum lags dominate over reverberation lags in the considered frequency range, the reverberation signal is still present.
Subtle reverberation features can be seen in Fig.~\ref{fig:pivoting_lags}, such as oxygen ($\sim 0.6-0.7$ keV)
and iron ($\sim 6-7$ keV) features. However, it would be  
extremely hard to distinguish these features once the telescope response is accounted for in this particular configuration of parameters.

\subsection{High density and ionisation variations}
\label{sec:high_dens_ion_var}
\begin{figure*}
	\includegraphics[width=2\columnwidth]{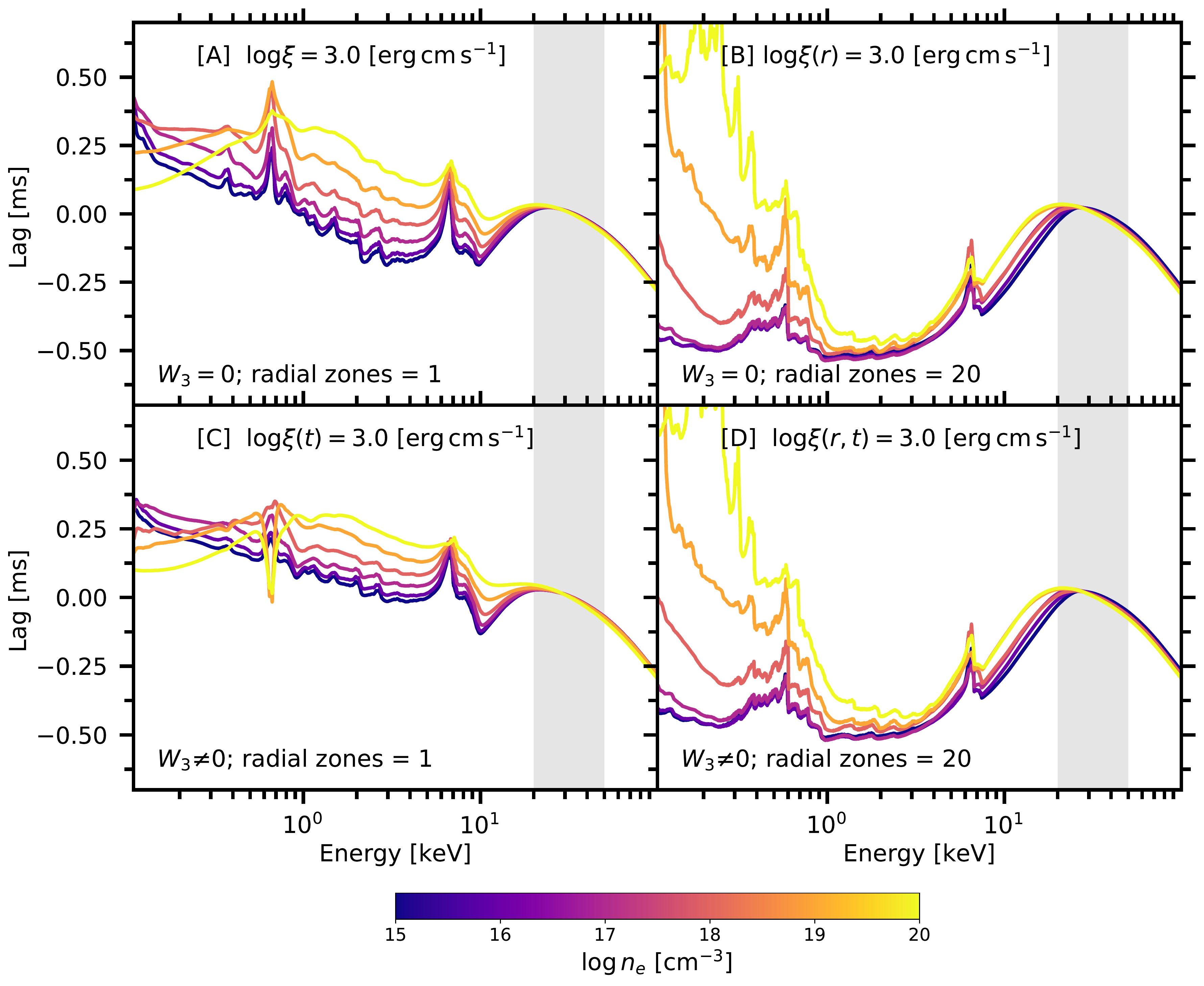}
        \caption{Lag energy spectra calculated by the \reltransDCp\ model in the
        frequency range $1-10$ Hz. All the parameters are specified in
        Table~\ref{tab:defaut_param} apart from the density which varies according to the different colors. The reference band of the cross-spectrum 
        calculation is between $20$ and $50$ keV (the grey area), thus all the spectra
        converge to zero lag at the Compton hump. 
        In the left panels (A and C) the ionisation is radially 
        constant over the entire disc (radial zones $= 1$), whereas in the right panels (B and D) 
        the disc is divided in $20$ zones (radial zones $= 20$) and the
        ionisation is calculated self-consistently. 
        Moreover, in the upper panels (A and B) the time variations of the ionisation 
        is neglected ($W_3 = 0$), whereas in the bottom panels (C and D) the ionisation is
        calculated accounting for the time variations of the illuminating radiation ($W_3 \neq 0$).}
    \label{fig:lag_density}
\end{figure*}
\begin{figure*}
	\includegraphics[width=\textwidth]{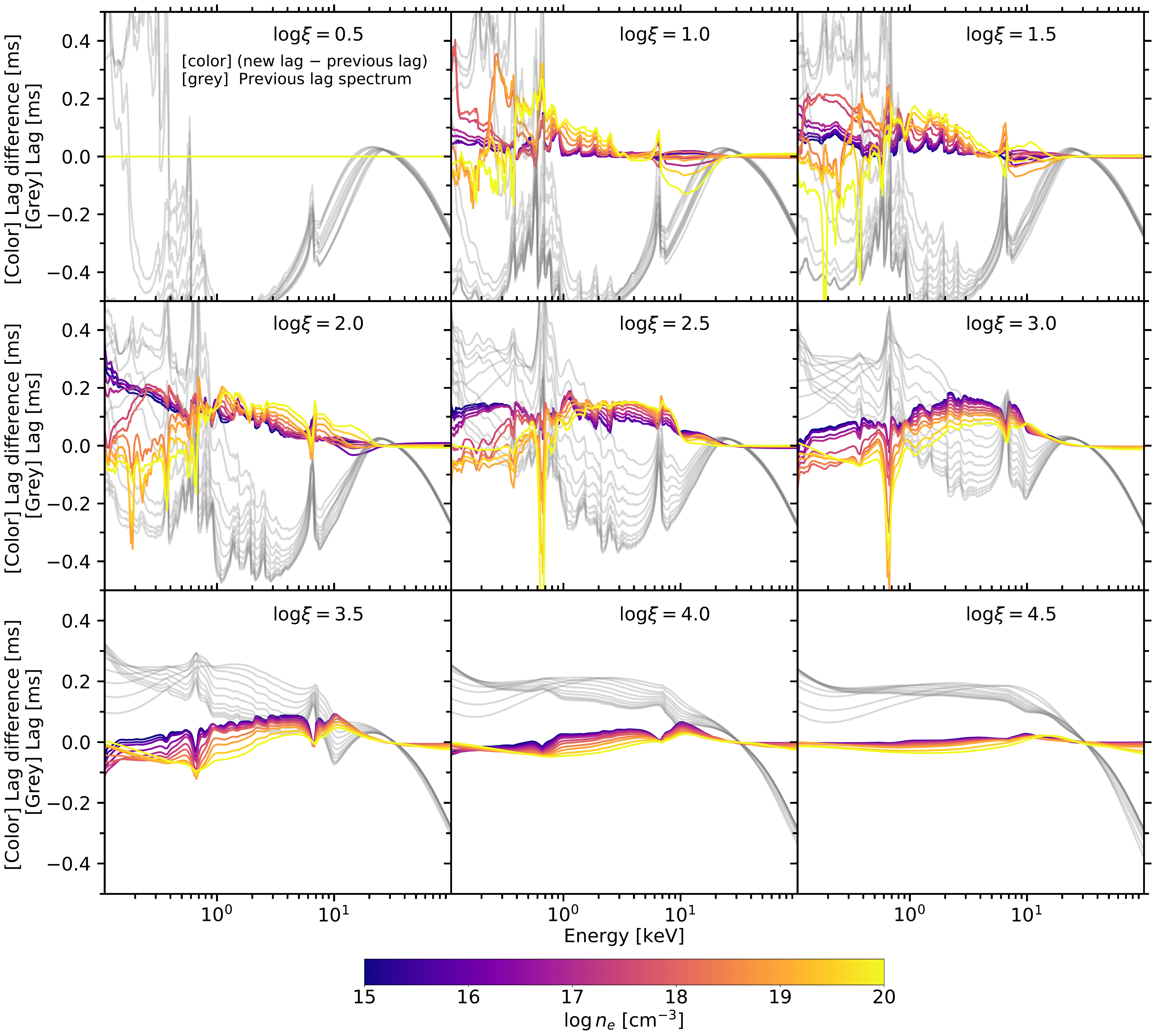}
    \caption{The colored lines represent the difference in 
    lag-energy spectrum between the version of 
    the \reltransDCp\ model where the time variations of the ionisation parameter 
    are neglected and the new version where they are accounted for by adding the 
    $W_3$ transfer function. The grey lines are the lag energy spectra of the old model
    version (neglecting time variation of the ionisation) shown as comparison to the difference
    between the old and the new version. Each panel shows the results for a range of
    of accretion-disc densities at a given ionisation parameter (in units of erg\,cm\,s$^{-1}$),
    as indicated. The rest of the model parameters were set to their default values (Table~\ref{tab:parameters}).
    The lag spectra were computed in the $1-10$\,Hz frequency range.}
    \label{fig:diff_lag_NOprofile}
\end{figure*}

\begin{figure*}
	\includegraphics[width=\textwidth]{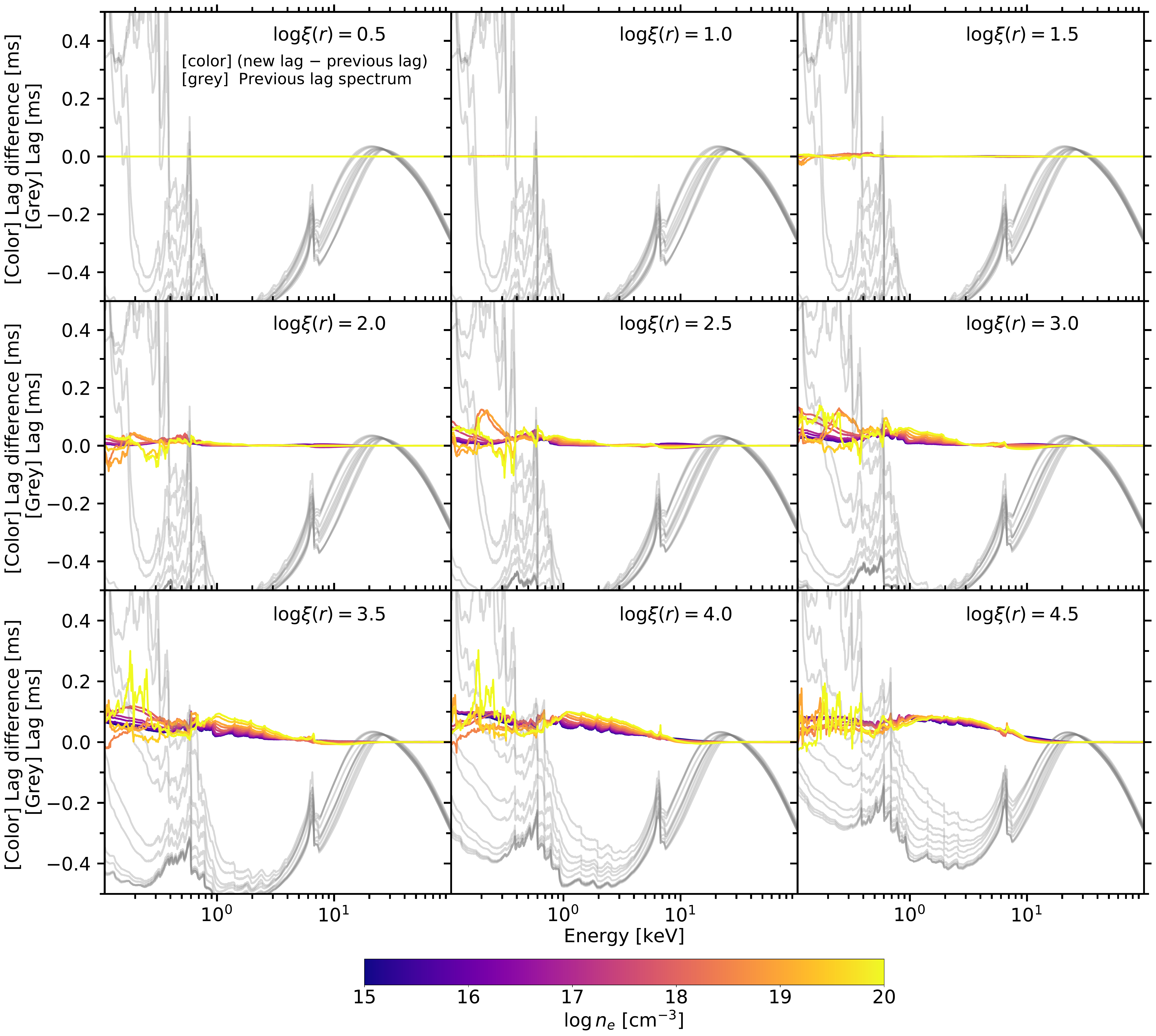}
    \caption{Same comparison between the old and the new version of the model as in Fig.~\ref{fig:diff_lag_NOprofile}, with the only difference 
    that the models account for a radial profile of the ionisation in the disc. The ionisation value of each panel is the peak value of the ionisation radial profile that is calculated 
    considering $20$ radial zones. The density is constant over these radial zones, 
    but it changes for each line as indicated by their color. The lag spectra are computed at the 
    frequency range of $1$ - $10$ Hz. The other the parameters of the models are equal to the default values (Table~\ref{tab:parameters}).}
    \label{fig:diff_lag_profile}
\end{figure*}

As in v1.0, \reltrans~v2.0 includes a radial ionisation profile, $\xi(r)$, by splitting the disc into a number of zones, each with a different value of $\xi$, and summing over the contribution from each of these zones. The number of zones is set by the environment variable \texttt{ION\_ZONES}. The simplest case of $\xi$ being independent of radius can therefore be recovered by setting \texttt{ION\_ZONES}=1. The $\xi(r)$ profile is calculated by computing the X-ray flux that crosses each disc annulus in the lamp-post geometry and dividing it by an assumed density profile $n_{\rm e}(r)$. There are two options for the assumed form of $n_{\rm e}(r)$: constant (environment variable \texttt{A\_DENSITY}=0) or the profile corresponding to zone A\footnote{$n_{\rm e} \propto r^{3/2} \left[1-(r_{\rm in}/r)^{1/2}\right]^{-2}$
} 
of the \cite{Shakura1973} disc model (\texttt{A\_DENSITY}=1). The ionisation parameter is normalised such that the maximum value of $\log\xi(r)$ is equal to the model parameter $\log\xi$.

Even though \reltrans~v1.0 included the option to calculate $\log\xi(r)$ 
by assuming a density profile, the value of the density was only used 
in the calculation of the ionisation radial profile.
The value of $n_{\rm e}$ used in the rest frame reflection spectrum 
was always fixed to 
$n_{\rm e}=10^{15}$\,cm$^{-3}$ (since this is hardwired into the \texttt{xillver}
and \texttt{xillverCp} tables). This inconsistency is now solved in v2.0, 
which uses the publicly available \xillverD\ and \xillverDCp\ tables that take
$n_{\rm e}$ as an input parameter in the range $n_{\rm e}=10^{15-20}$\,cm$^{-3}$ \citep{Garcia2016}. 
In the new model, each radial zone can have a different value of $\xi(r)$ \textit{and}
of $n_{\rm e}(r)$ to input into the calculation of the rest frame reflection spectrum. 
The model parameter $n_{\rm e}$ specifies the smallest value that $n_{\rm e}(r)$ reaches across the entire disc extent
(when the profile is constant it sets the density value in the whole disc).
In addition, v2.0 also accounts for $\xi$ varying in time, driven by the variations
in the illuminating flux. This feature is implemented by the inclusion of an extra 
transfer function $W_3$ (see Section~\ref{sec:formalism}), and therefore can be
switched off by artificially setting $W_3=0$.

Fig.~\ref{fig:lag_density} shows the effects on the lag-energy spectrum of changing disc density and of considering radial and time variations of the ionisation parameter.  
It is not always straightforward to understand these effects because of the 
non-linear dependence of the lag-energy spectrum on the flux-energy spectrum. 
Therefore, we assume a constant radial profile for the disc density ({\tt A\_DENSITY} $=0$) throughout the rest of this paper. 
We remind the reader of the possibility of including a radial density profile in the model ({\tt A\_DENSITY} $=1$). 
For the rest of this section, we use \reltransDCp\ with the default parameters (Table~\ref{tab:defaut_param}) and $\phi_{AB} = \gamma = 0$, meaning that there are no continuum lags because the shape of the continuum spectrum is not varying. The reference band is set at the Compton hump ($20-50$ keV) where the lag is zero for all the densities. 
In our simulations the ratio between the continuum flux and the reflection flux is self-consistently 
calculated assuming the properties of the system. The height of the lamp-post ($6\,R_{\rm g}$), the spin of the black hole (maximum spinning) and the inner radius (ISCO) of the accretion disc are the model parameters that set the geometry and determine the reflection fraction as defined in \citet{Dauser2016}. The model can deviate from a static lamp-post configuration by varying the boost parameter which can either enhance ($1/\mathcal{B} > 1$) or diminish ($1/\mathcal{B} < 1$) the reflection flux.
However, in our simulations and later in our fit to the data the boost parameter is always fixed to $1$ (static lamp-post geometry). 

In Fig.~\ref{fig:lag_density} we consider six values of $n_{\rm e}$ (as labelled in the color bar), and each panel explores a different scenario for the ionisation parameter:

\noindent A) Ionisation $\xi$ is constant over time and radius ($W_3=0$ and {\tt ION\_ZONES} $=1$); 

\noindent B) Ionisation $\xi(r)$ is constant over time and varies over radius ($W_3=0$ and {\tt ION\_ZONES} $=20$);

\noindent C) Ionisation $\xi(t)$ is constant over radius and varies over time ($W_3\neq0$ and {\tt ION\_ZONES} $=1$);

\noindent D) Ionisation $\xi(r,t)$ varies over time and radius ($W_3\neq0$ and {\tt ION\_ZONES} $=20$). 
\vspace{0.5pt}

The main effect of increasing the density on the lag-energy spectrum is the enhancement of the 
lag at low energies (below $1$ keV) where the reflected flux increases.
This effect is more evident when we account for the radial profile of the 
ionisation parameter (panels B and D). This happens because the ionisation as a function of radius
has a steep negative gradient due to the illuminating flux 
(the density is radially constant in the disc), hence the outer
regions of the disc contribute with less ionised reflection. 
These spectra have a more prominent soft excess, 
which is enhanced for higher densities (\citealt{Garcia2016}). 
It is clear that the shape of the lag-energy spectrum depends closely on 
the shape of the flux-energy spectrum, since the dilution of 
the lags is set by the relative amplitude 
of the reflection with respect to the continuum emission (see discussion in \citealt{Cackett2014}).
This soft excess feature in the lag-energy spectrum is much more 
subtle for the configurations with $\xi$ independent of radius (panels A and C).
In fact, for $n_{\rm e} \gtrsim 10^{19}$\,cm$^{-3}$ a break appears at $E\sim 0.5$ keV. This is caused by the shift of the soft 
excess to higher energies, which is also seen in the flux-energy spectrum (\citealt{Garcia2016}).

Arguably the most evident feature is the drastic change of the oxygen 
line (O~\textsc{viii}  Ly$\alpha$) around $0.65$ keV (panel C).
The `inversion' of this emission line at high densities 
occurs only when we consider the time variations 
of the ionisation parameter and no radial profile. 
The oxygen emission line is extremely sensitive 
to the ionisation parameter and its relative ionic fraction
peaks around \logXi{\, = \,2} (\citealt{Kallman2020}). 
Therefore, even though the time variations of the ionisation parameter are small
($\delta\log\xi \sim 0.04$) they 
are enough to over-ionise the oxygen line for a range of $n_{\rm e}$ values 
such that increasing the illuminating flux reduces the strength of 
the line rather than increasing it.
The same effect also happens for all of the other emission lines, 
but never as clearly as for oxygen. 
For some lines, this is simply because they are less prominent 
in the spectrum and for others it is because the peak relative ionic fraction 
occurs at a higher value of the ionisation than \logXi{\,=\,3}.
We note that if we consider a radial ionisation profile (panel D), 
the contribution from the outer parts of the disc with less ionised spectra
prevents this effect from becoming visible in the final spectrum. 
In this case the total spectrum includes contributions from a range of 
ionization parameter values, all less than \logXi{\,=\,3}. 
Since the oxygen line does not over-ionise in the outer disc, the contributions of the outer radii dilute the overall effect in the lag spectrum, leading to the oxygen feature
simply being weaker at high $n_{\rm e}$ rather than inverting.

We see from Fig.~\ref{fig:lag_density} that the introduction of the ionisation parameter
time variations can have a large effect on the lag-energy spectrum if 
we consider $\xi$ to be independent of radius (panels A vs panel C), 
but only a subtle effect when we employ a radial $\xi$ profile (panel B vs panel D). 
We further investigate this by plotting the difference between the lag-energy spectrum 
calculated accounting for the ionisation time-variations ($W_3 \neq 0$) 
and that calculated neglecting them ($W_3 = 0$).
Fig.~\ref{fig:diff_lag_NOprofile} shows this time lag difference 
for nine values of $\log\xi$ (colored lines), under the assumption 
that $\xi$ is independent of radius. 
The grey lines are the time lags calculated without ionisation time-variations,
shown to give an idea of the importance of the difference between the two calculations
(without dividing by the lag, which can be zero).
Once again the model parameters are specified in Table~\ref{tab:defaut_param}, 
in particular the height of the lamp-post source ($6\,R_{\rm g}$), spin of the black hole (maximally spinning) and inner radius (ISCO)
of the disc are fixed to keep the same reflection fraction in the model.
For near neutral gas (\logXi{\,=\,0.5}), the time variations of the ionisation parameter 
do not produce any notable additional lags. This is because 
the atomic features do not change significantly over a small range of ionisation when the gas is close to being neutral. 
Therefore, the flux-energy spectrum does not change shape. 
When the gas is more highly ionised ($1.0\leq$ \logXi{} $\leq 3.5$) the lag
produced by the ionisation variation becomes significant, 
especially at higher densities (yellow lines). In this ionisation range
the atomic features are sensitive even to small variations of the ionisation parameter.
The right central panel (labelled `$\log\xi=3.0$') corresponds to 
the difference between panels C and A of Fig. \ref{fig:lag_density}. 
The drastic difference at the oxygen line is again clear, and 
it is now possible to see that the iron line feature becomes significantly 
weaker when ionisation variations are considered. 
Finally when the gas starts to become completely ionised (\logXi{} $\geq 4.0$) 
the variations of the illuminating flux do not change the ionisation 
state of the gas and the lag contribution drops, pushing the difference 
close to zero as in the case of the neutral gas. 
We note that, for intermediate values of $\log\xi$, the amplitude 
of the lag difference (colored lines) can be comparable at low energies, 
if not even larger than, the lag spectrum without ionisation 
time variations (grey lines). 
In these cases, the introduction of ionisation time-variations 
has therefore changed the lags by $100\%$ or more.
However, this extreme difference is likely to be an overestimate, 
since in this scenario the ionisation level of the entire disc 
is described by a single ionisation value.

Fig.~\ref{fig:diff_lag_profile} is the same as Fig.~\ref{fig:diff_lag_NOprofile},
except now a radial ionisation profile is included in the model
(i.e. the right central panel in Fig.~\ref{fig:diff_lag_profile}
corresponds to the difference between panels D and B of Fig. \ref{fig:lag_density}),
which we consider to be more realistic. 
We see that the additional lag introduced by ionisation time-variations 
is in general much smaller than before. 
This is because the model parameter $\log\xi$ sets the peak value of $\log\xi(r)$, 
thus even when the ionisation peak indicates a completely ionised gas, 
there is still a significant contribution from the less ionised spectra radiated 
from further out in the disc. 
When the gas is neutral the ionisation time-variations do not contribute to the lag,
therefore the lags due to the ionisation time-variations are less prominent 
when we account for the radial ionisation profile. 
We do see that the additional lag contributed by ionisation variations 
increases as the peak ionisation is increased. 
This occurs because increasing the peak ionisation leads to a smaller fraction of the disc being neutral. 
However, overall the change in the lag spectrum is small, with the change never exceeding $80\%$ and typically being $\lesssim 10\%$.

\begin{figure*}
	\includegraphics[width=2\columnwidth]{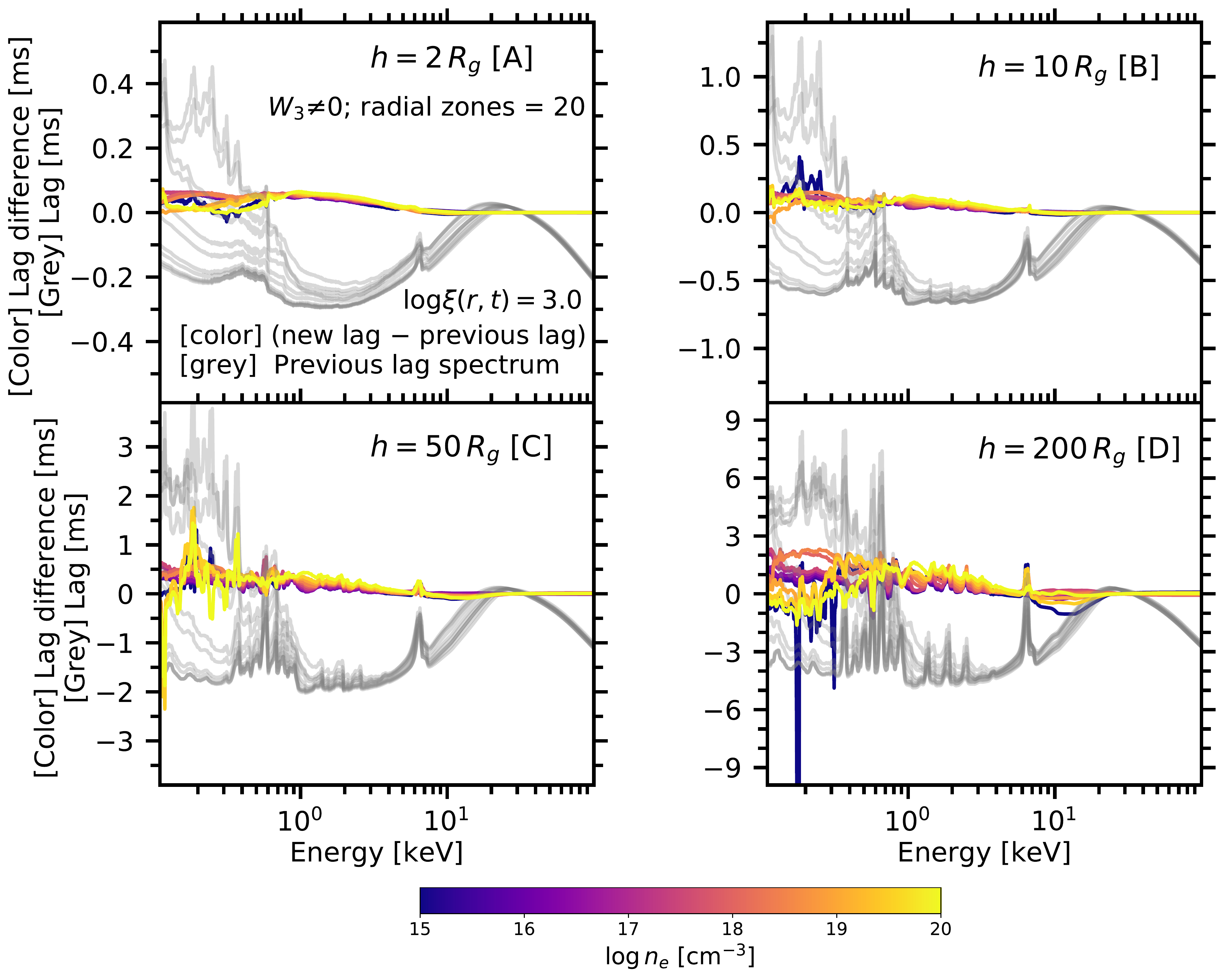}
        \caption{
        Same comparison between the old and the new version of the model as in Fig.~\ref{fig:diff_lag_profile}, with the  difference that the ionisation parameter is fixed to \logXi{} $= 3$ in every panel and that the height of the source varies among the panels. The ionisation value of each panel is the peak value of the ionisation radial profile that is calculated 
    considering $20$ radial zones. The density is constant over these radial zones, 
    but it changes for each line as indicated by their color. The lag spectra are computed at the 
    frequency range of $1$ - $10$ Hz. The other the parameters of the models are equal to the default values (Table~\ref{tab:parameters}).}
    \label{fig:diff_lag_profile_height}
\end{figure*}

Finally we show how the ionisation time-variations change the lag energy spectrum 
for different heights of the lamp-post source. Changing the source height affects both the
light crossing time of the photons and the reflection fraction in the model,
therefore the amplitude and energy dependence of the lags are also different.
Fig.~\ref{fig:diff_lag_profile_height} shows the same concept of Fig.~\ref{fig:diff_lag_NOprofile} and \ref{fig:diff_lag_profile}, though the peak ionisation is fixed to \logXi{} $= 3$ and the height of the source 
is different among the panels ($2$, $10$, $50$ and $200$ $R_{\rm g}$, from A to D respectively). 
We note that the scale on the y-axis changes in the panels, since the distance between the primary source and the disc changes quite significantly. 
It is worth noting that the amplitude of the colored lines, representing difference between the lag-energy spectrum 
calculated accounting for the ionisation time-variations 
and the lag-energy spectrum calculated neglecting them, is increasing compared to the corresponding 
grey lines, which represent the lag-energy spectra with no ionisation time-variations.
This means that the effects of the ionisation time-variations are more relevant for larger values 
of the source height. 
Of particular interest it is to look at this difference at the iron line energy range.
Panels A and B show a negligible amplitude of the colored lines compared to the grey lines.
The colored lines start to be relevant in panel C ($50\, R_{\rm g}$) where it is of the same order of magnitude of the grey lines, and they are important in panel D ($200\, R_{\rm g}$) where they modify the amplitude of the iron line lag by more than $50\% $.

\section{The case of MAXI~J1820+070}
\label{sec:maxi1820}
\begin{figure}
	\includegraphics[width=\columnwidth]{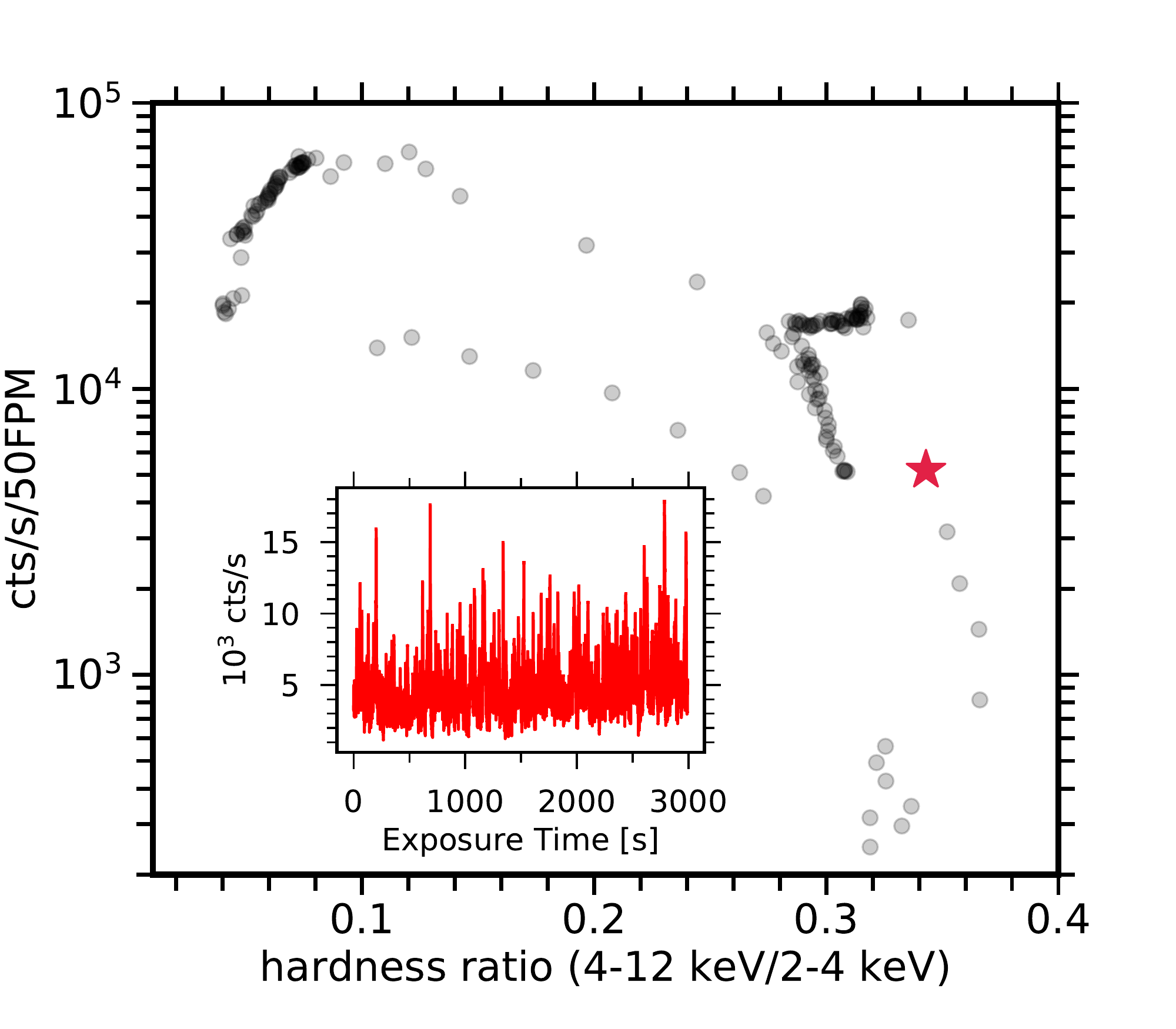}
    \caption{Hardness intensity diagram for the 2018 outburst of MAXI~J1820+070 observed with \nicer . The red star is the \nicer\ observation analysed in this work. The inset plot shows the light curve of the red star observation without the gaps. }
    \label{fig:HID}
\end{figure}

We test \reltrans ~2.0 on the recently discovered BHB 
MAXI~J1820+070. The system went into outburst in 2018 and it was one of
brightest black hole transients ever observed (\citealt{Kawamuro2018}). The \textit{Neutron Star Interior Composition Interior Explorer}
(\nicer; \citealt{Gendreau2016}) managed to track the source over the entire outburst from the 
typical rise in the hard state, through the transition to the soft state, 
and across the transition back to the hard state before the source went back to quiescence. 
Fig.~\ref{fig:HID} shows the full outburst of MAXI~J1820+070 where 
the circles are all the \nicer\ observations of the source. 
An iron line reverberation feature was detected in the lag vs energy spectrum during 
the hard state (\citealt{Kara2019}), which makes this source 
a very good candidate to test our new model. 
We consider an observation during the rise of the hard state 
that was not analysed by \citet{Kara2019}. 
In Fig.~\ref{fig:HID} the red star identifies the observation that we  
use in this work and the inset shows its full light curve.
Even though the source showed quasi periodic oscillations (QPOs)  
during its outburst, none were detected in this 
observation. 
Since the iron line profile has been observed to vary systematically 
with QPO phase \citep{Ingram2016,Ingram2017} in a manner that is very likely 
due to changes in the accretion flow geometry, 
observations such as this one with no QPO are the most suitable to 
analyse with \reltrans, which assumes a constant accretion geometry.
We fit the lag-energy spectrum simultaneously in five ranges of the Fourier frequency 
and investigate the different contributions to the total lags in the model. 

\subsection{Data reduction}
We consider the \nicer\ observation taken on 2018, March 16 (ObsID 1200120105). 
We use data-analysis software NICERDAS version 2019-05-21\_V006 and CALDB version xti20200722 with the energy scale (gain) version ``optmv10" to analyse the observation.
We applied a series of filtering criteria: 
1) neglecting the noisy detectors  FPMs 34 and 14; 
2) selecting only events that are not flaged as ``overshoot" or 
``undershoot" resets (EVENT\_FLAGS=bxxxx00), or forced triggers (EVENT\_FLAGS=bx1x000); 
3) applying the ``trumpet" filter to clean the observation out 
from known background events (\citealt{Bogdanov2019a});
4) considering only events when the pointing offset is less than $54\arcsec$, the pointing direction is 
more than $40^\circ$ away from the bright Earth limb, 
more than $30^\circ$ away from the dark Earth limb, 
outside the South Atlantic Anomaly (SAA).
The \texttt{FTOOL} barycenter correction \texttt{barycorr} 
is applied to the cleaned events.
We consider the cross-spectrum between two light curves in the energy bands 
$0.5$ - $1$ keV and  $1.5$ - $3$ keV and we compute the time lag as a function 
of Fourier frequency. 
We also consider $24$ smaller energy range light curves between 
$0.5$ keV and $10$ keV and calculate the 
cross-spectrum between each of those and the full energy range light curve
as a reference band. 
The subject light curve is always subtracted from the reference band light curve, 
thereby ensuring they are statistically independent.
In order to fit the lag spectrum, \reltrans\ requires the Redistribution 
Response Matrix (RMF) and the Auxiliary Response Matrix (ARF) in its calculations.
For this we use the RMF version ``rmf6s" and the ARF version ``consim135p", which are 
both part of the CALDB version xti20200722.

\subsection{Analysis of the Lag Spectra}

\begin{figure*}
	\includegraphics[width=\textwidth]{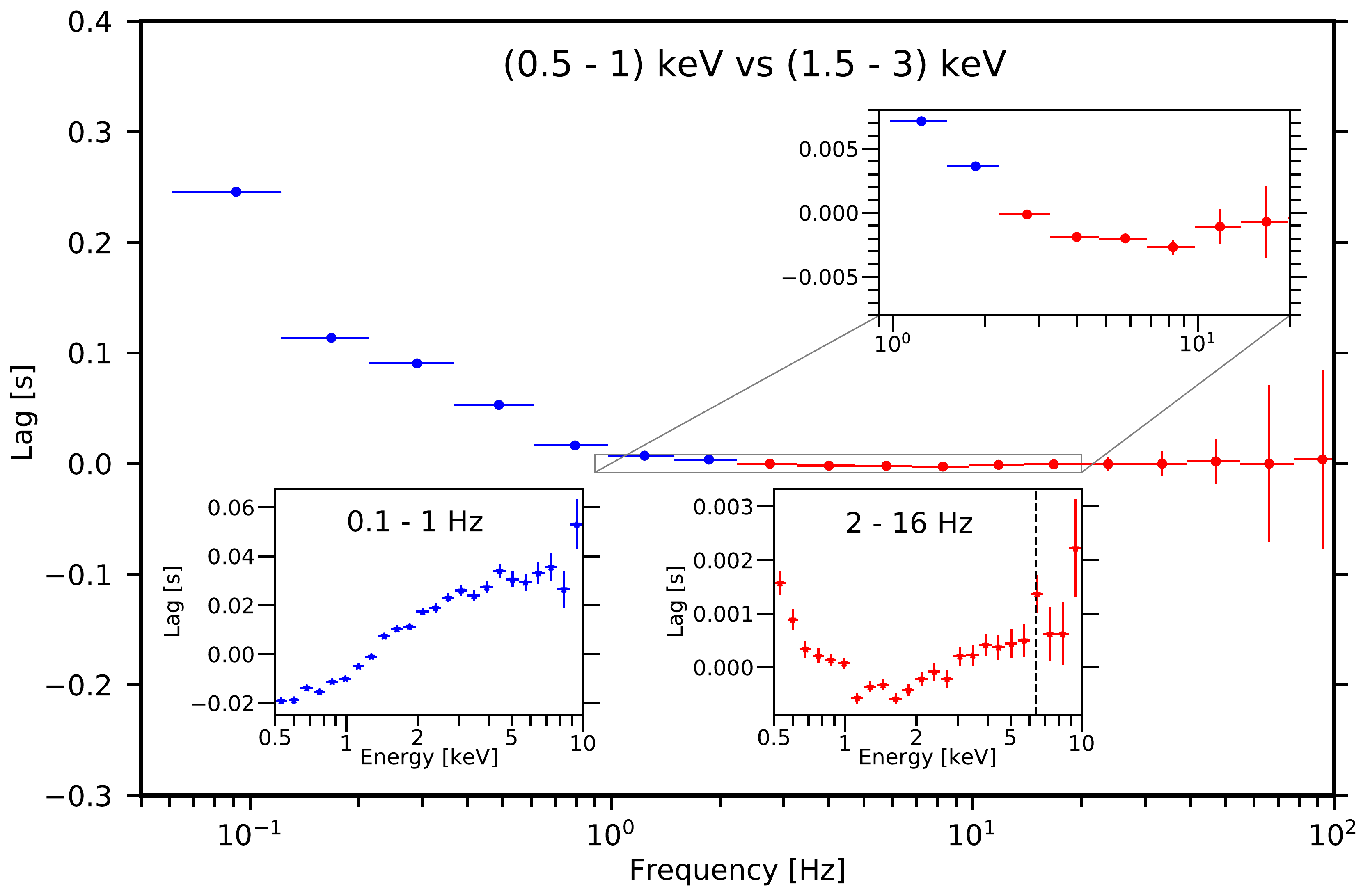}
    \caption{Lag vs frequency spectrum of MAXI~J1820+070 during the rise of the hard state. The cross spectrum was calculated between the $0.5$ - $1$ keV light curve and the $1.5$ - $3$ keV light curve. The blue points are positive hard continuum lags and red points are negative soft reverberation lags. The Top inset zoom in on the negative lags, the color code is the same as the main plot. The two bottom insets show the lags as a function of energy in the hard continuum lag frequency range (left panel, blue points) and in the soft reverberation lag frequency range (right panel, red points).The vertical dashed line marks the position of the iron line at $6.4$ keV.}
    \label{fig:lags_nicer}
\end{figure*}
We first consider the lag between the $0.5$ - $1$ keV
and $1.5$ - $3$ keV light curves. 
Fig.~\ref{fig:lags_nicer} shows the lag as a function of Fourier frequency. 
The positive hard lags (blue points) dominate at 
low frequencies (up to $\sim 1$ Hz), whereas at higher frequencies 
the lags turn negative as a sign that the reverberation lags dominate (red points).
The top right inset of Fig.~\ref{fig:lags_nicer} magnifies the negative lags 
showing that the amplitude is of the order of a few milliseconds.  
The intrinsic difference between these two types of 
lag becomes clear by looking at their energy dependence. 
The two lower insets of Fig.~\ref{fig:lags_nicer} show 
the energy dependence of the lags
in the two different variability timescales. 
At long timescales (low Fourier frequencies, $0.1$ - $1$ Hz) the lags increase monotonically 
as a function of energy, whereas at short 
timescales (high frequencies, $2$ - $16$ Hz) the lag spectrum presents a peculiar shape that resembles 
the flux energy reflection spectrum with a hint of a feature
around the iron line energy range (indicated by the dashed vertical line).

Since the \reltrans~2.0 model accounts for both continuum and 
reverberation lags, it has the capability of fitting the lag energy
spectrum at any timescale.
This allows us to follow the evolution of the different 
contributions to the total lags with Fourier frequency
and thus reduce the degeneracy between the model parameters.
We fit simultaneously five lag energy spectra with \reltransDCp\
accounting for time variations
of the ionisation parameter and a self-consistent 
radial profile of the ionisation in the accretion disc. 
The latter is calculated considering $20$ radial zones 
in the disc and keeping the electron density constant among them
({\tt ION\_ZONES} $=20$ and {\tt A\_DENSITY} $=0$). 
The peak value of the radial ionisation profile and 
the density of the disc are free parameters in the model.
All the free parameters are specified in Table~\ref{tab:parameters} and Table~\ref{tab:pivoting_parameters}. 
The former includes the parameters that are tied for all the frequency ranges, and the latter includes the continuum variation parameters, which are allowed to vary in each frequency range.  
All the parameters that are not listed in these tables are fixed in our fit. 
The iron abundance and the hydrogen column density of
Galactic absorption are always fixed to the Solar value and 
to $10^{21}$\,cm$^{-2}$, respectively 
(see e.g. \citealt{Uttley2018ATel, Shaw2021} 
as references for low absorption column). 
We fixed the spin and the mass of the black hole to $0.5$ (\citealt{Buisson2019}) which implies that the innermost stable circular orbit (ISCO) is at $4.2\,R_{\rm g}$ and $10 M_{\odot}$, respectively.
Furthermore, we fixed the boost parameter to $1$. 
This choice forces the model to be consistent with a static, isotropic 
radiating and point-like corona. 
The reflection fraction is calculated self-consistently from the geometry of the system that is determined by the height of the corona and the inner radius. These parameters are free to vary in the fit, although they are tied together among the frequency ranges. For whatever set of parameters the data prefers, the model always calculates internally the reflection fraction.

\begin{figure*}
	\includegraphics[width=\textwidth]{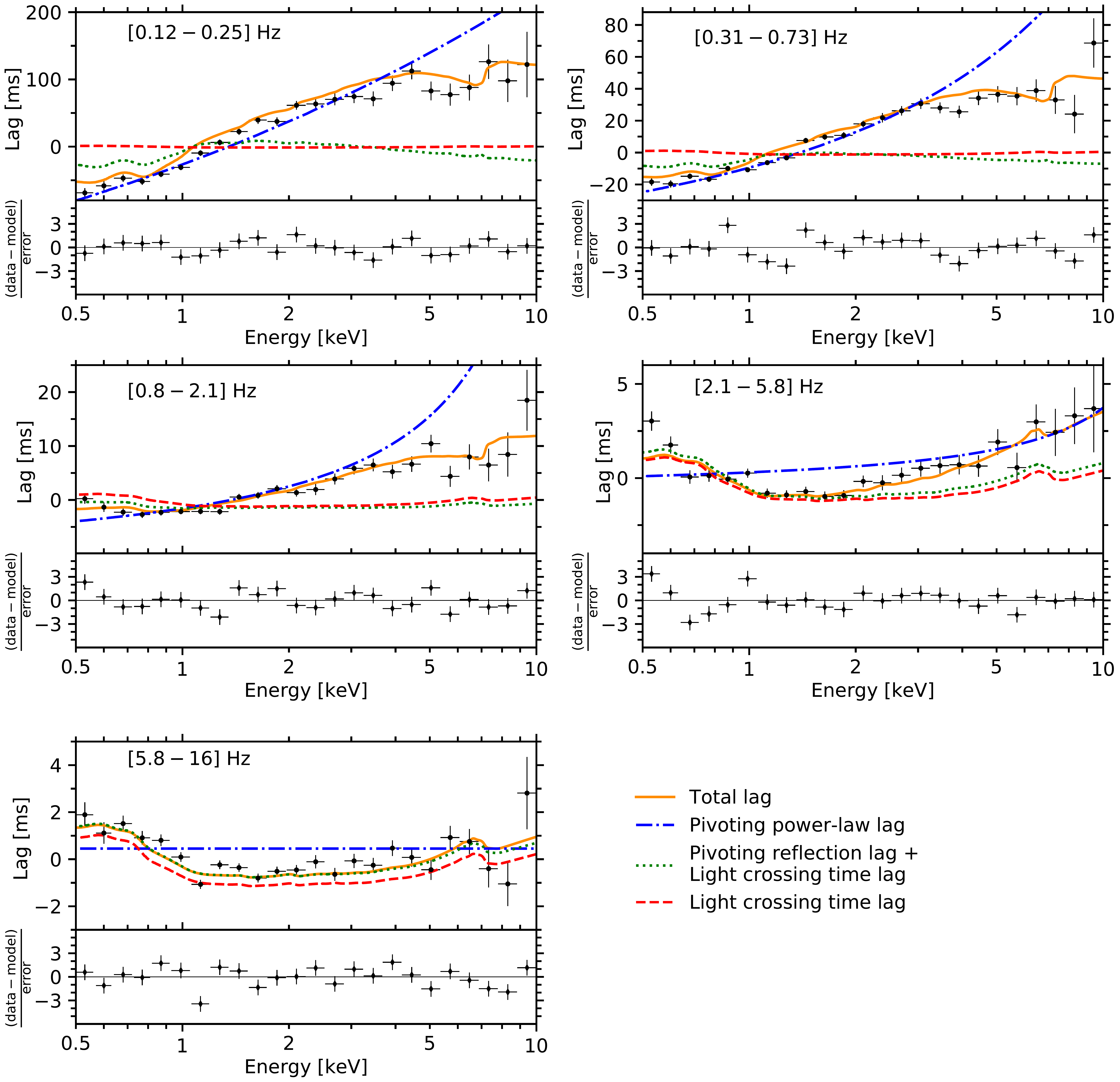}
    \caption{\reltransDCp\ fit of the lag energy spectra in $5$ frequency
    ranges with Model 1a. The iron abundance and the boosting parameters are fixed to 
    $1$ and the mass parameters to $10 M_{\odot}$.
    Each panel show the best fit total lag model 
    (solid purple lines) and the contributions to the total lag. 
    The blue dashed dotted lines show the the lag due to the 
    illuminating pivoting continuum, the dotted green lines show the lag 
    due to the different light crossing time between the continuum and reflected emission together with the lag caused by the pivoting reflection,
    and the dashed red lines neglect the the pivoting reflection showing the
    contribution of just the light crossing time. The bottom panels show the
    residuals of the best fit to the binned model in terms of sigmas with error bars of size one.}
    \label{fig:best_fit}
\end{figure*}
Fig.~\ref{fig:best_fit} shows the lag vs energy spectra in the five 
frequency ranges together with the best fit \reltransDCp\ 
model (solid orange lines).
The rest of the lines in the plots represent different contributions to 
the total lag calculated by the model. 
The dashed dotted (blue) lines show the lags due to only the variation of 
the power-law index (pivoting power-law) without any transfer function involved in the calculation. Thus, in this case all the transfer functions are set
to zero and the lag-energy spectrum always shows a log linear shape with no reflection features. 
These lags dominate at relatively low frequencies, where they set the amplitude of the total lag. At higher frequencies, however, these power-law intrinsic lags are negligible. Nevertheless, the fit shows that these lags are not able to predict the correct shape of the lag-energy spectrum, not even at the lowest frequencies. Therefore other sources of lag need to contribute.

The dotted (green) lines show the lags caused by the reflection emission. These lags are due to both the light-crossing time lag and time variations in the hardness of the reflection spectrum in response to hardness variations of the illuminating spectrum.
Finally, the dashed (red) lines show only the light-crossing time lag (without accounting for the response of the reflection spectrum to variations in the slope of the illuminating spectrum).
These lags dominate at relatively high frequencies. 
Nonetheless, their contributions at relatively low frequencies
are essential to reproduce a lag-energy spectrum that deviates from being strictly log linear, as is observed.
Looking at the different contributions it is clear that the total lag is not the sum of the pivoting power-law lags (dot-dashed line) and the reflection lags (dotted line),
instead these contributions have to be correctly combined in the 
calculation of the cross-spectrum. 
In fact, it is \textit{not} correct to add the lags produced by different effects, \textit{the correct procedure is to sum the complex cross-spectrum components} and then calculate the lag spectrum (see Appendix~A in M18). 

The residuals in the bottom panels of each plot of Fig.~\ref{fig:best_fit}
show neither any evident broad structure nor any large systematic offset, indicating that the model succeeds in
fitting simultaneously all the frequency ranges and therefore the different types of lags. 
We note, however, that a few data points diverge from the 
model with more than $3\sigma$ significance. These are either narrow structures around $1$ keV that the model struggles to reproduce or 
a few points at low energies where the model underestimates the amplitude of the lag. One clear case is the top right panel where the $>3\sigma$ residuals are easy to spot. 
This could be due to over-simplistic modelling of the absorption features that could affect the dilution effect in the spectrum.
It is worth noting that in the iron line region the residuals are within $3\sigma$ level showing good agreement between the model and the data.
It is remarkable how the model follows the overall shape of the lag energy spectra among the different frequency ranges: at low frequencies reproducing the dips in the data corresponding to the reflection emission lines $\sim0.6$\,keV, $\sim0.8$\,keV and the iron line; and at high frequency the positive Fe line feature and the strong soft excess.
\citet{Wilkins2016} previously predicted some of these features in the lag-energy spectrum, such as the iron line dip at low frequencies. Those authors compared their simulations with lag-energy spectra of AGN in order to constrain the geometry of the corona. However, this current work is the first time that an iterative fitting routine could be applied to constrain these reverberation features in the lag energy spectrum of a stellar-mass black hole.

The best fit $\chi^2$ and the model parameters are shown in Table~\ref{tab:parameters}. 
The best fit parameters of the continuum variability are listed in 
Table~\ref{tab:pivoting_parameters}. These continuum variability parameters are free to vary over frequency due to the phenomenological nature of the continuum lag in our model. The only a priori constraint on these parameters was to limit $\phi_{\rm AB}$ to negative values, which forces the intrinsic continuum variability to produce only hard lags. We note that the inner radius is pegged to its lower limit, indicating that the accretion disc extends to the 
ISCO, thus suggesting no truncation.
We also note that the electron 
density of the accretion disc is pegged to its upper limit of $n_{\rm e}=10^{20}$\,cm$^{-3}$,
which supports the prediction from a simple $\alpha$-disc model (\citealt{Shakura1973}) that accretion discs in 
stellar-mass black holes are much denser than the $n_{\rm e}=10^{15}$\,cm$^{-3}$ value traditionally assumed for AGN.
The $\phi_A$ parameter is always smaller than $0.01$ except for the last frequency range. This means that 
the correction to the lag spectra to account for poor calibration is almost negligible. This is expected, since the response matrix is well calibrated in the $0.5-10$ keV energy band that we used for our reference band light curve.
The $\phi_{AB}$ and $\gamma$ parameters are well constrained and they are compatible with $0$ for the highest frequency range. This shows that, in this frequency range, the pivoting power-law lags do not contribute significantly to the fit. 

\begin{figure}
	\includegraphics[width=\columnwidth]{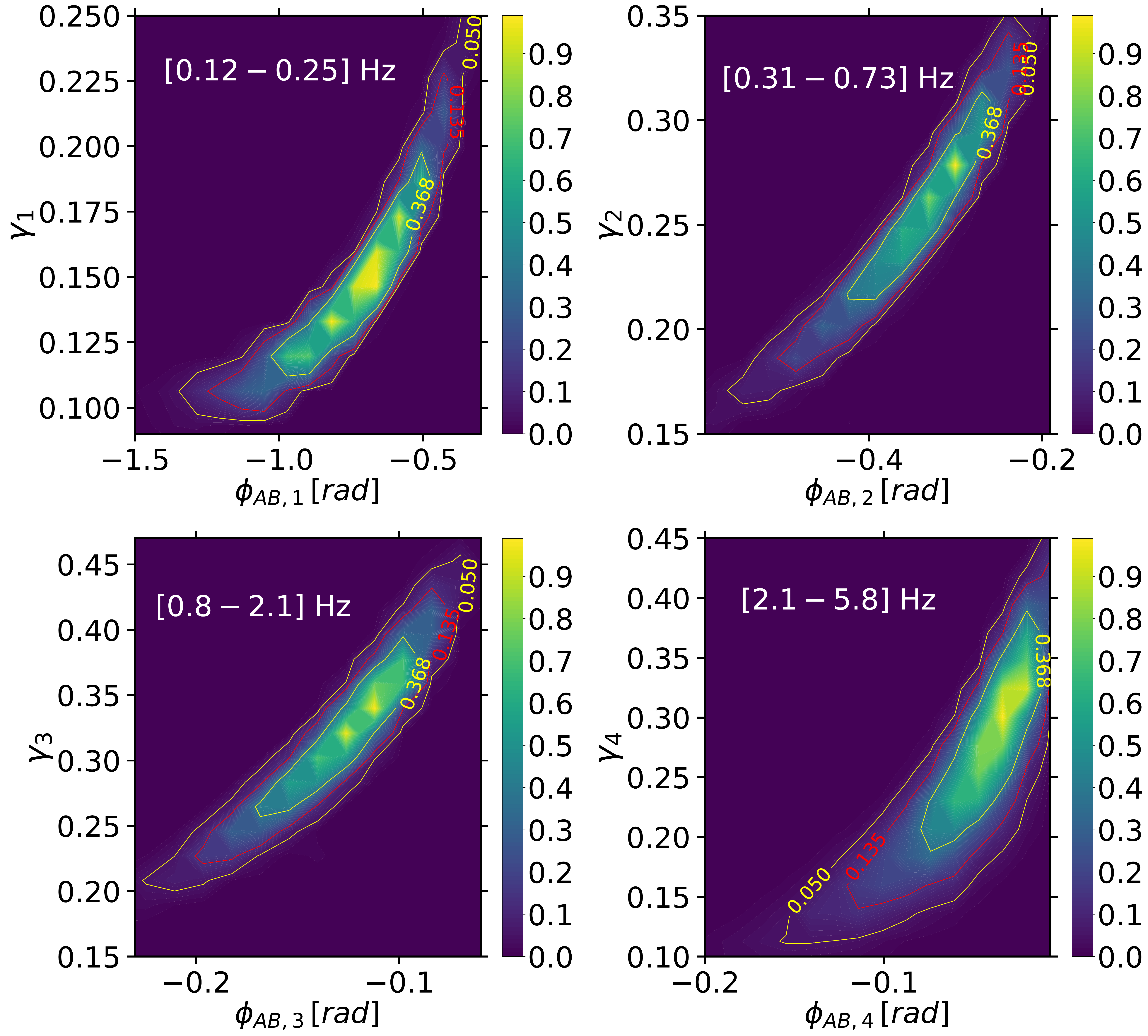}
    \caption{Contour plots of the pivoting power-law parameters resulting from the MCMC analisys of our fit to the \maxi . The color bars indicate the probability distribution of the two parameters.}
    \label{fig:contour_pivoting}
\end{figure}

We performed a Markov Chain Monte Carlo (MCMC) analysis to probe the parameter space of our fit and investigate possible correlations among the parameters. The details of the analysis are reported in Appendix~\ref{app:MCMC} as well as all the possible combinations of the 2D contour plots among the free parameters. 
It is worth noting that there is no degeneracy between the continuum parameters and the reverberation parameters (see Fig.~\ref{fig:corner_plot}), although some parameters are pegged to their limits such as the inner radius (pegged at ISCO) and the density (pegged at $10^{20}$\,cm$^{-3}$).
Having no clear degeneracy between continuum and reflection parameters ensures that the model can distinguish between the two types of lags. This is a direct consequence of forcing $\phi_{\rm AB}$ to be negative in the fit, thus preventing the continuum lags from contributing to the soft lags.
Fig.~\ref{fig:contour_pivoting} shows a sub-set of contour plots focusing on the pivoting parameters. 
We note the clear correlation among $\phi_{AB}$ and $\gamma$ in all the frequency ranges shown. The parameters are labeled with numbers starting from the lowest frequency range considered. The last frequency range has been excluded since the two parameters are compatible with zero.
This confirms that the model requires a better signal-to-noise at higher energy resolution of the lag-energy spectrum to break the degeneracy between the continuum parameters.

\begin{table}
\renewcommand{\arraystretch}{1.5}
\caption{Model free parameters for our \reltransDCp\ best fit of MAXI~J1820+070 \nicer\ obsID 1200120105. Spin, $n_{\rm H}$, Mass and Boost parameters are fixed to $0.5$ (ISCO $= \,4.2\,R_{\rm g}$), $10^{21}$\,cm$^{-2}$, $10\,M_{\odot}$ and $1$ respectively. 
These free parameters have the same values for the five frequency ranges we consider in the fit. 
The pivoting parameters are specified in Table~\ref{tab:pivoting_parameters} since they are free to vary in each frequency range.}

\begin{tabular}{  p{1.5cm} | m{1.5cm} }
  \hline
\multicolumn{1}{c} {Parameter} &
\multicolumn{1}{c} { \makecell{Best fit values}}  \\
  \hline
$h$ \,[$R_{\rm g}$] & $30^{+9}_{-3}$   \\

Incl  [deg]         &    $44^{+8}_{-8}$       \\

$ r_{\rm in}$ [ISCO] & $< 1.5$ \\ 

$\Gamma$             & $1.793^{+0.003}_{-0.031}$     \\

$\log \xi$         & $1.4^{+0.1}_{-0.1}$  \\

$\log n_{\rm e}$         & $> 19.7$ \\ 

$\chi^2/\mathrm{d.o.f.}$      & $ 166/120 $ \\
  \end{tabular}
    \label{tab:parameters}
 \end{table}

\begin{table}
\renewcommand{\arraystretch}{1.5}
\caption{Best fit pivoting parameters for all the five frequency ranges.}
\begin{tabular}{ l|c|c|c}
  \hline
Frequency & $\phi_{A}$ & $\phi_{AB}$ & $\gamma $  \\
  \hline

$[0.12 - 0.25]$   &  $-0.017^{+0.003}_{-0.003}$  &  $-0.7^{+0.1}_{-0.2}$ &  $0.15 ^{+0.06}_{-0.05}$ \\
$[0.31 - 0.73]$ &  $-0.018^{+0.003}_{-0.003} $ & $-0.3^{+0.1}_{-0.1}$  & $0.28^{+0.04}_{-0.04}$  \\
$[0.8 - 2.1]$ &  $ -0.008^{+0.003}_{-0.003} $   & $-0.10^{+0.03}_{-0.05} $ & $0.37^{+0.08}_{-0.10} $ \\
$[2.1 - 5.8]$ &   $0.009^{+0.003}_{-0.003} $ & $-0.03^{+0.02}_{-0.04}$ &  $ 0.3^{+0.1}_{-0.1}  $ \\
$[5.8 - 16]$ & $0.031^{+0.008}_{-0.009}$   & $>-10^{-9}$& $0.09^{+0.1}_{-0.1}$ \\
  \end{tabular}

    \label{tab:pivoting_parameters}
 \end{table}

\begin{table}
\renewcommand{\arraystretch}{1.5}
\caption{Default parameters of the \reltransDCp\ model. }
\begin{tabular}{| l | l | l | c |}
  \hline
 & Parameters and description & \xspec\ name & default value \\
  \hline
  \hline
 1.&  \makecell[l]{\,$h$ \,[$R_{\rm g}$]  \\ coronal height }    & \,$h$                  & $6$     \\
  \hline
 2.&   \makecell[l]{\,$a$  \\  black hole spin}           & \,$a$                    & $0.998$ \\
  \hline
 3.&   \makecell[l]{\,$i$  \, [deg]  \\ Disc inclination}       & \,Incl                       & $30$    \\
  \hline
 4.&   \makecell[l]{\,$ r_{\rm in}$ \,[ISCO] \\ Disc inner radius} & \,$ r_{\rm in}$   & $-1$    \\
  \hline
 5.&   \makecell[l]{\,$ r_{\rm out}$ \,[$R_{\rm g}$] \\ Disc outer radius}&  \,$ r_{\rm out}$  & $1000$  \\
  \hline
 6.&   \makecell[l]{\,$ z $ \\ Cosmological redshift}  & \,$ z $& $0.0$   \\
  \hline
 7.&   \makecell[l]{\,$\Gamma$ \\ power-law spectral index}    & \,$\Gamma$         & $2.0$   \\
  \hline
 8.&   \makecell[l]{\,$\log [\xi/$erg\,cm\,s$^{-1}]$    \\ Disc ionisation (Ionisation peak \\ if radial profile is defined)}              & \,$\log \xi$     & $3.0$   \\
  \hline
 9.&   \makecell[l]{\,A$_{\rm Fe}$ \\ Iron abundance}                   & \,A$_{\rm Fe}$ & $1$     \\
  \hline
10.&   \makecell[l]{$\log [n_{\rm e}$/cm$^{-3}$] \\ Disc density (minimum density \\if radial profile is defined)}             &  $\log n_{\rm e}$  & $15$    \\
  \hline
11.&   \makecell[l]{kTe  \,[keV] \\ Coronal electron temperature}   &  kTe  & $60$    \\
  \hline
12.&   \makecell[l]{$n_{\rm H} $ \,[$10^{22}\,{\rm cm}^{-2}$] \\ TBabs column density} &  $n_{\rm H}$  & $0.0$   \\
  \hline
13.&   \makecell[l]{$1/\mathcal{B} $    \\ Reflection boost with respect \\to the power-law emission} & Boost  & $1$   \\
  \hline
14.&   \makecell[l]{Mass  \,[$M_{\odot}$] \\ Black hole mass}               & Mass & $10$    \\
  \hline
15.&  \makecell[l]{ f$_{\rm min}$  \,[Hz] \\ Minimum of the \\ frequency range}            & f$_{\rm min}$ & $1$     \\
  \hline 
16.&   \makecell[l]{f$_{\rm max}$  \,[Hz] \\ Maximum of the \\ frequency range}           & f $_{\rm max}$  & $10$    \\
  \hline
17.&   \makecell[l]{ReIm \\ Model output}      & ReIm & $-1$    \\
  \hline
18.&   \makecell[l]{$\phi_{A}$ \\ Normalisation phase}   & \texttt{phiA} & $0.0$    \\
  \hline
19.&   \makecell[l]{$\phi_{AB}$ \\ Phase difference between \\ the spectral index and the \\normalisation variability  }                            & \texttt{phiAB} & $0.0$    \\
  \hline
20.&   \makecell[l]{$\gamma$ \\Amplitude ratio between \\the spectral index and the \\normalisation variability  } & \texttt{g}& $0.0$    \\
  \hline

  \end{tabular}
    \label{tab:defaut_param}
 \end{table}

\section{Discussion}
\label{sec:discussion}

We presented the new release of the \reltrans\ model (\reltrans~2.0).
This model aims to fit the the hard lags at low Fourier frequency and the soft lags at high Fourier frequency of accreting black holes. We studied the hard lags parameters considering the new mathematical formalism and the new features introduced in the reflection lags such as ionisation time-variations and the effects of accounting for high densities (above $10^{15}$\,cm$^{-3}$) in the accretion disc. 

\subsection{Physical motivation for the pivoting lags}
We have investigated the two model parameters $\phi_{AB}$ and $\gamma$ that are 
respectively the amplitude ratio and the phase difference between the power-law spectral index variability ($\Gamma(t)$) and power-law normalization variability ($A(t)$). We now relate these two parameters that control the 
pivoting variability of the power-law radiation
to the \textit{rms} variability observed from accreting black hole systems. 
The mathematical assumption of the model is that most of the power-law time variability originates from the normalisation variations, nonetheless, a small portion of the total \textit{rms} is due to time variations of the power-law spectral index. 
From a physical perspective, if the origin of the non-thermal emission is 
attributed to Comptonization of seed disk photons in a hot corona (\citealt{Haardt1993, Zdziarski2003}), then the variability
of $\Gamma$ can be related to the variability 
of the temperature $T_{\rm e}$ and the optical depth $\tau_{\rm e}$ 
in said corona as  
\begin{equation}
\label{eq:gamma_tau_temp}
    \Gamma = -\frac{1}{2} + \sqrt{\frac{9}{4} + \frac{1}{\theta_{\rm e} \tau_{\rm e}(1+ \tau_{\rm e}/3)}}
\end{equation}
where $\theta_{\rm e}  = kT_{\rm e }/ m_{\rm e } c^2$, and $m_{\rm e }c^2 = 511$ keV is the electron rest mass energy (\citealt{Lightman1987}).
Eq.~\ref{eq:gamma_tau_temp} is estimated using empirical relations and it is valid for $\tau > 1$.
Thus, to produce time variations of $\Gamma$, either the temperature or the optical depth of the corona need to vary with time (or both). Changes in the temperature are caused by variations in the balance between Compton heating and cooling rates, which depend on the shape
of the radiation field and on the electron density. Changes in the optical depth can also be caused by variations in the electron 
density, or by a modification in the geometrical extension of the corona.
The amplitude of $\Gamma$ variations in our model is related to the parameter $\gamma$. 
As we showed in Section~\ref{sec:pivoting_parameters} assuming $\gamma = 0.5$ and $\sim 10\%$ \textit{rms} variability of the power-law normalisation, this corresponds to a $\sim 2\%$ fractional \textit{rms} variation in $\Gamma$. Adopting mean values of $\overline{kT}_{\rm e}=30$\,keV and $\overline{\tau}_{\rm e} = 3$ into eq.~\ref{eq:gamma_tau_temp}, we calculate that this $2\%$ fractional \textit{rms} variability in $\Gamma$ can be produced by $3\%$ \textit{rms} variability in both $kT_{\rm e}$ and $\tau_{\rm e}$.
A Taylor expansion to first order of eq.~\ref{eq:gamma_tau_temp} around small time variations in both $\theta_{\rm e}$ and  $\tau_{\rm e}$
shows that the \textit{rms} variability in $\Gamma$ decreases when the mean values of the coronal temperature and optical depth increase.

\subsection{High density in the reverberation lags}
The new \reltrans~2.0 model is designed to use the high 
density reflection tables of the \xillverD\ and \xillverDCp\ models
(\citealt{Garcia2016}), which allow us to reach more realistic 
values of the electron density in the accretion disc. 
The reverberation lags follow the shape of the flux-energy spectrum, 
and the effect of high densities is to enhance the reflected flux at soft energies (below $1$\,keV). Therefore considering high densities in the lag-energy spectrum will result in an enhancement
 of the lags at soft energies. 
This effect is more evident when the ionisation of the disc is self-consistently calculated by accounting for a radial dependent illuminating flux. In this case, 
the outer part of the disc is less ionised and thus its contribution to the soft energies is more prominent (e.g., see panel~D in Fig.~\ref{fig:lag_density}). 
It is worth noting that soft lags at low energies 
have been observed in BHBs (\citealt{Uttley2011, DeMarco2017}).
They have been interpreted as signatures of the  
different light crossing time between the direct photons and reprocessed illuminating photons 
that are re-emitted as quasi-thermal radiation (\citealt{Uttley2011}). 
We note that by considering high-density reflection models we have included the quasi-thermal component of the re-processed emission. Therefore, this process is now correctly accounted for in our model. 

We note that the thermalisation timescale, 
the characteristic time for the photons to lose their energy through scattering, 
has always been considered negligible when compared to the light crossing timescale. 
The same applies to the photoelectric absorption and recombination timescales 
which depend on the density, temperature and ionisation of the accretion disc (see discussion in \citealt{Wilkins2020}). 
Even though \reltrans\ fits well the lag energy spectra, we note that 
for the highest frequency ranges the model misses a few data points at the lowest energies (Fig.~\ref{fig:best_fit}).
Perhaps despite the high density of the disc, at these high frequencies the thermalisation timescales
are long enough such that they could significantly contribute to the observed lags.
Investigating this is beyond the scope of this paper and will be a topic of future work. 

\subsection{Fit of \maxi }
For a test case, we have chosen a hard state observation 
of the BHB MAXI~J1820+070 performed by \nicer .
We fit the lag as a function of energy in five frequency ranges simultaneously, 
probing the hard and soft lags in the lag-frequency spectrum. 
These are clearly two different types of lags, one related to the continuum emission
and the other to the reflected emission due to reverberation. 
However, the two processes are not completely separated: the reflected
emission depends strongly on both the illuminating flux and its variability.
With our model we show the importance of accounting for both these 
processes simultaneously when fitting observational data. 
The fit of MAXI~J1820+070 shows that the reflection lags contribute significantly at the relatively low frequencies, since the contribution of these lags is the only way to shape the lag-energy spectrum
differently than the log linear behaviour described by the hard lags.
In the first panel of Fig.~\ref{fig:best_fit} it is clear how the overall lag (orange solid line) drops by almost $50\%$ from the pivoting lag (blue dash-dotted line)
when we account for the contribution of the reflection lags (green dotted line). Moreover, the dips at low energies ($\sim0.6$\,keV and $0.8$\,keV) and at the Fe line ($\sim6$\,keV) are signatures of the reverberation lags that dilute the dominant hard continuum lags. 
These reflection lags are caused not only by the light crossing time delay. There is an additional source of lags produced by variations in the hardness and ionisation of the reflected radiation, which are caused by variations in the illuminating spectrum.
By correctly accounting for all these lags in the complex cross-spectrum, \reltrans\ is able to match the irregular shape of the lag-energy spectra at all timescales. 

We note that in our fit the lags at the highest frequency do not show 
a strong iron line feature, neither when we combine the frequencies between $2-16$ Hz, nor 
when we look at those frequencies in separate ranges.
However, there is a hint of an iron line when we combine the frequencies, with marginal significance due to the relatively low signal-to-noise ratio of this observation. 
The lack of the iron line signal in the observation analyzed here means that the reverberation lags are constrained by the low energy part of the lag-energy spectrum.
In a separate publication we present detailed fits of the \reltrans\ model to higher luminosity observations
of MAXI~J1820+070 during the same outburst (\citealt{Wang2021}). 

Even though we fit this observation as a proof of principle and we do not consider the flux energy spectrum, it is interesting to compare our results with previous works. \citet{Buisson2019} fitted the flux energy spectrum of the \nustar\ observations, the first one is during the rise of the hard state. They used a dual-lamp-post corona model to fit the flux energy spectrum and constrained the heights at $3\, R_{\rm g}$ and $100\, R_{\rm g}$. Our value of the height ($30_{-3}^{+9} \, R_{\rm g}$) falls in the middle of the two lamp-posts. The dual-lamp-post geometry approximates the vertical extension of the corona, thus it could be expected that our single lamp-post model used in the lag spectrum fitting constrains the height of the source roughly in the middle of the two vertical edges. 
However, we caution the reader about this interpretation, since the fit of the flux energy spectrum and the fit of lag-energy spectrum done with \reltrans\ during the hard-to-soft transition (\citealt{Wang2021}) lead to the opposite trends of the coronal height\footnote{\citet{Wang2021} used a version of \reltrans\ with high density, but without ionisation time-variations.}. 
The analysis of \citet{Wang2021} shows that the coronal height derived from the flux-energy spectrum fit decreases from $\sim 25\, R_{\rm g}$ to $\sim 9\, R_{\rm g}$, whereas the coronal height from the lag fit increases from $\sim 40\, R_{\rm g}$ to $> 300 \, R_{\rm g}$. These results suggest that the observed reverberation lags come mainly from the upper part of a vertically extended corona, instead of from the the entire structure of the corona. 
Finally, we note that the soft lags are $\sim3$ ms (upper inset panel of Fig.~\ref{fig:lags_nicer}) during the rise of the hard state. These lags are larger than what is reported by \citet{Kara2019} who analysed the bright hard state. However, the decreasing trend of the soft-lag amplitude during the bright hard state seems to hold also during the rising of the outburst, since the amplitude of the first bright hard state observation is $\sim1$ ms (\citealt{Kara2019}). This would fit the picture of the shrinking corona described in \citet{Kara2019}.

\section{Conclusions}
We present a new release of our model for spectral and timing 
variability in accreting black holes: \reltrans ~2.0. 
We address variability in both the direct and reflected emission, and 
their energy dependence, by modelling the complex cross-spectrum as a function 
of energy at any Fourier frequency range. 
In this work we have described the new formalism and how the new features
affect the shape of the lag vs energy spectrum, in particular.
We investigated the new features applied to the model and here we list the main results: 

\begin{itemize}
    \item The low Fourier frequency variability can be modeled 
    through time variations of both the normalisation and the hardness of the continuum radiation. In particular the low frequency lags can successfully be described by the pivoting power-law mechanism.  $2\%$ \textit{rms} variability in the continuum spectral index is sufficient to produce the observed time lags. This variability comes from the reasonable assumption that the temperature and the optical depth of the corona vary at $3\%$ \textit{rms}. 
    \item Flux variations of the coronal emission cause time variations of the ionisation level in the accretion disc. The variations in the reflected spectral shape are  an additional source of time lags that can change features in the lag-energy spectrum. Depending on the accretion disc radial profile that we consider, the ionisation time-variations can change the spectrum even more than $100\%$ 
    \item We found that using high density accretion disc models when calculating the reflected emission change the shape of the lag-energy spectrum as compared to lower density models. We considered densities from $10^{15}$\,cm$^{-3}$ to $10^{20}$\,cm$^{-3}$ and we found an enhancement of the reverberation lags below $1$ keV when increasing the density. 
    This effect is due to the quasi-thermal component of the reflected emission becoming more prominent at higher densities. 
    \item We fitted the lag-energy spectrum in multiple Fourier frequency ranges in one \nicer\ observation of \maxi\ during the rise of the hard state as a proof of principle. 
    The best fit shows how the different lag-components of the model contribute at different timescales. 
    At low Fourier frequencies (below $2$\,Hz) the pivoting lags dominate the lag-energy spectrum, but the reverberation lags contribute to shape the lag-energy spectrum. At high Fourier frequencies (above $2$\,Hz) the reverberation lags due to light crossing time delays are dominant. The model well reproduces the shape of the lag spectra, with sensible values of the parameters. 
\end{itemize}

\section*{Acknowledgements}
The authors would like to acknowledge the anonymous referee for the very helpful comments and suggestions.
GM, JW, EK and JAG acknowledge support from NASA grant 80NSSC17K0515. AI acknowledges support from the Royal Society. MK thanks the support from NWO Spinoza. JAG thanks support from the Alexander von Humboldt Foundation. This work was partially supported under NASA contract No. NNG08FD60C.

\section*{Data Availability}
The source code for the model described in this paper will soon be made publicly available in the form of an update to the reltrans model package (\url{https://adingram.bitbucket.io/reltrans.html}).
The \nicer\ observation is publicly available at the HEASOFT webpage.
 



\bibliographystyle{mnras}
\bibliography{library_2020}



\appendix

\section{MCMC analysis}
\label{app:MCMC}
\begin{figure*}
    	\includegraphics[width=\textwidth]{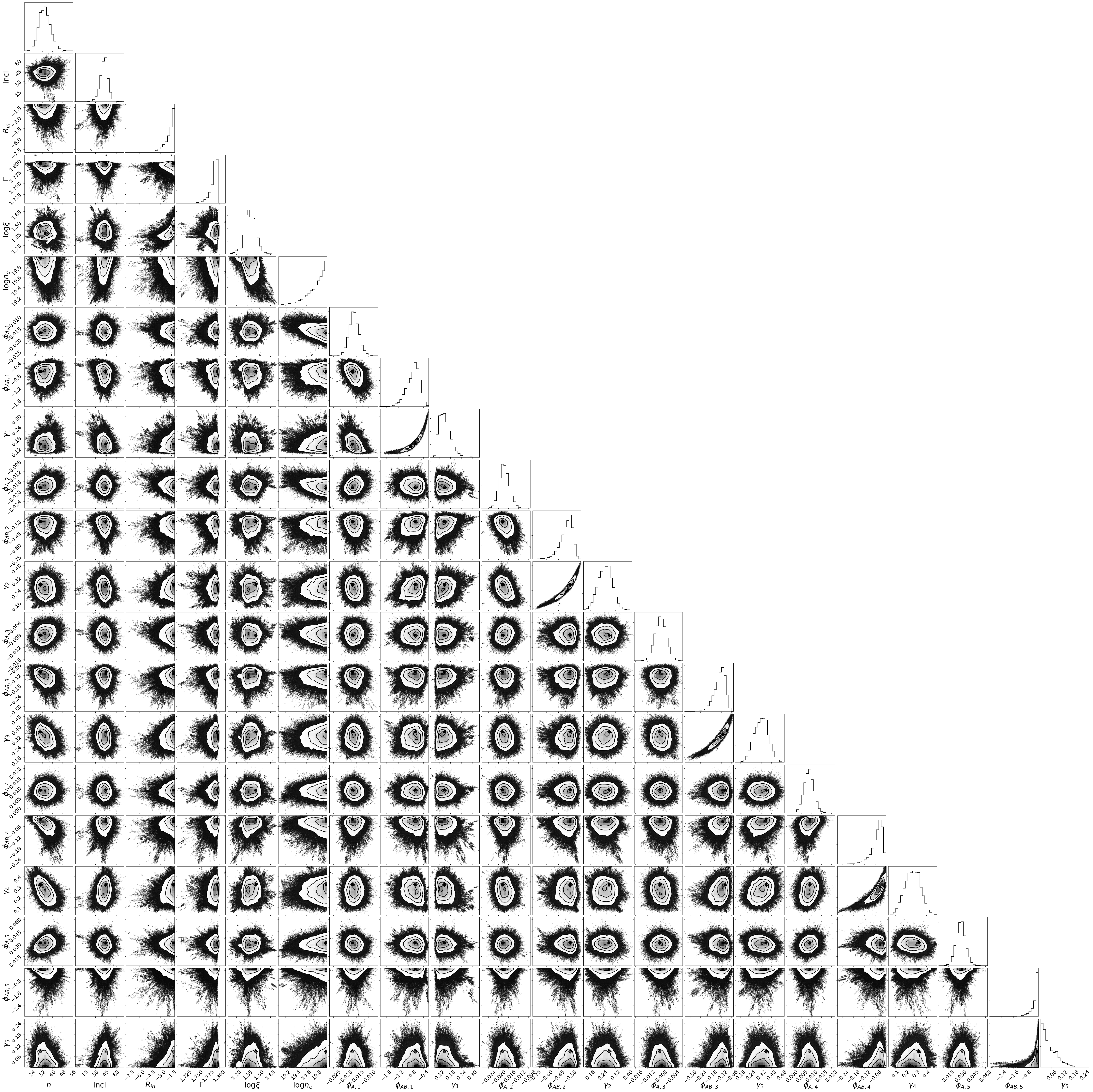}
    \caption{Corner plots for all the free model parameters in our best fit. The subscript to the pivoting parameters correspond to the frequency ranges from lowest to highest.}
    \label{fig:corner_plot}
\end{figure*}
We performed MCMC analysis using the Goodman–Weare algorithm. We run four chains, using the \xspec\ routine \texttt{chain}. The total length is $10^{6}$ steps and we use 200 walkers. All the walkers start from the best fit parameters and we set the burn-in period to $5000$ steps.
Fig.~\ref{fig:corner_plot} shows the corner plot for all the model 
parameters free to vary during the fit. The $\phi_{A}$, $\phi_{AB}$ and $\gamma$ parameters are labeled with numbers from the lowest to the highest frequency range considered.


\label{lastpage}
\end{document}
